\documentclass[10pt,letterpaper]{article}
\usepackage[top=1.5in,left=1.6in,right=1.6in,footskip=1.5in]{geometry}

\usepackage{amsmath,amssymb}

\usepackage{changepage}

\usepackage[utf8x]{inputenc}

\usepackage{textcomp,marvosym}

\usepackage{cite}

\usepackage{nameref}
\usepackage[hyperfootnotes=true,colorlinks]{hyperref}
    	\hypersetup{colorlinks,%
    	citecolor=blue,%
    	filecolor=green,%
    	linkcolor=blue,%
    	urlcolor=blue,%
   	}

\usepackage[right]{lineno}

\usepackage[table]{xcolor}

\usepackage{array}

\usepackage{verbatim}

\newcolumntype{+}{!{\vrule width 2pt}}

\newlength\savedwidth

\usepackage[aboveskip=1pt,labelfont=bf,labelsep=period,justification=raggedright,singlelinecheck=off]{caption}

\makeatletter
\renewcommand{\@biblabel}[1]{\quad#1.}
\makeatother

\date{}

\usepackage{lastpage,fancyhdr,graphicx}
\usepackage{epstopdf}

\usepackage{nameref}
\usepackage{times}
\usepackage{longtable}

\newcommand{\half}[0]{\frac{1}{2}}

\newcommand{\Sp}{\textrm{S}}
\newcommand{\Dp}{\textrm{D}}
\newcommand{\SSs}{\Sp\Sp}
\newcommand{\SSz}{\SSs_0}

\newcommand{\SSpi}{\SSs_\pi}
\newcommand{\I}{\textrm{I}}
\newcommand{\DD}{\Dp\Dp}
\newcommand{\DS}{\Dp\Sp}
\newcommand{\SD}{\Sp\Dp}

\newcommand{\SSS}{\Sp\Sp\Sp}

\newcommand{\SDS}{\Sp\Dp\Sp}
\newcommand{\DSD}{\Dp\Sp\Dp}

\newcommand{\ud}{\textrm{d}}
\newcommand{\udi}{\,\ud}

\newcommand{\Tor}{\mathbb{T}}
\newcommand{\R}{\mathbb{R}}
\newcommand{\Z}{\mathbb{Z}}

\newcommand{\sset}[1]{\left\lbrace #1\right\rbrace}

\newcommand{\abs}[1]{\left|#1\right|}

\DeclareMathOperator{\Var}{Var}
\DeclareMathOperator{\Img}{Im}
\DeclareMathOperator{\Rep}{Re}

\renewcommand{\Im}[1]{\Img\!\left(#1\right)}
\renewcommand{\Re}[1]{\Rep\!\left(#1\right)}

\newcommand{\maxdim}{N}
\newcommand{\maxpop}{M}

\newcommand{\C}{\mathbb{C}}

\newcommand{\ks}{k_s}

\newcommand{\as}{\alpha_{s}}

\newcommand{\vth}{\vartheta}
\newcommand{\omb}{\widehat{\omega}}
\newcommand{\etb}{\widehat{\eta}}

\newcommand{\OP}{Z}

\newcommand{\be}{\begin{equation}}
\newcommand{\ee}{\end{equation}}

\newcommand{\omf}{\omega_\mathrm{f}}

\newcommand{\alphacoeff}{a}

\newcommand{\WS}{(\hyperref[eq:WSrh]{WS})\ }

\newcommand{\Torn}{\Tor^{\maxdim}}

\newcommand{\psiCM}[2]{\psi^{(#1)}_#2}
\newcommand{\BP}{z}

\newcommand{\avg}[1]{\left\langle#1\right\rangle}
\newcommand{\gt}{\mathrm{g}}
\newcommand{\Vrev}{V^\textrm{rev}}

\newcommand{\N}{\mathbb{N}}
\newcommand{\Pn}{P^{(n)}}

\newcommand{\ts}{\tau_\mathrm{syn}}

\newcommand{\tn}{\mathrm{tn}}
\newcommand{\kgj}{\kappa^{\textsc{gj}}}
\newcommand{\kg}{\kappa^{\gt}}

\newcommand{\scl}{1}

\usepackage{etoolbox}%
\apptocmd{\thebibliography}{\small}{}{}%

\begin{document}

\begin{flushleft}
{\Large
\textbf\newline{Understanding the dynamics of biological and neural oscillator networks through exact mean-field reductions: a review} 
}
\newline
\\
Christian Bick\textsuperscript{1--5},
Marc Goodfellow\textsuperscript{2,3,6,7},
Carlo R.~Laing\textsuperscript{8},
Erik A.~Martens\textsuperscript{9--11}
\\
\bigskip
\textbf{1} Centre for Systems, Dynamics, and Control, University of Exeter, Exeter, UK
\\
\textbf{2} Department of Mathematics, University of Exeter, Exeter, UK
\\
\textbf{3} EPSRC Centre for Predictive Modelling in Healthcare, University of Exeter, Exeter, UK
\\
\textbf{4} Mathematical Institute, University of Oxford, Oxford, UK
\\
\textbf{5} Institute for Advanced Study, Technische Universität München, Garching, Germany
\\
\textbf{6} Living Systems Institute, University of Exeter, Exeter, United Kingdom
\\
\textbf{7} Wellcome Trust Centre for Biomedical Modelling and Analysis, University of Exeter, Exeter, UK
\\
\textbf{8} School of Natural and Computational Sciences, Massey University, Auckland, New Zealand
\\
\textbf{9} Department of Applied Mathematics and Computer Science, Technical University of Denmark, Kgs.~Lyngby, Denmark
\\
\textbf{10} Department of Biomedical Science, University of Copenhagen, Copenhagen N, Denmark
\\
\textbf{11} Centre for Translational Neuroscience, University of Copenhagen, Copenhagen N, Denmark
\bigskip

%
%






\end{flushleft}
\section*{Abstract}
Many biological and neural systems can be seen as networks of interacting periodic processes. Importantly, their functionality, i.e., whether these networks can perform their function or not, depends on the emerging collective dynamics of the network.
Synchrony of oscillations is one of the most prominent examples of such collective behavior and has been associated both with function and dysfunction. 
Understanding how network structure and interactions, as well as the microscopic properties of individual units, shape the emerging collective dynamics is critical to find factors that lead to malfunction. However, many biological systems such as the brain consist of a large number of dynamical units. Hence, their analysis has either relied on simplified heuristic models on a coarse scale, or the analysis comes at a huge computational cost. Here we review recently introduced approaches,  known as the Ott--Antonsen and Watanabe--Strogatz reductions, allowing one to simplify the analysis by bridging small and large scales. 
Thus, reduced model equations are obtained that exactly describe the collective dynamics for each subpopulation in the oscillator network via few collective variables only.
The resulting equations are next-generation models: Rather than being heuristic, they exactly link microscopic and macroscopic descriptions and therefore accurately capture microscopic properties of the underlying system. At the same time, they are sufficiently simple to analyze without great computational effort. In the last decade, these reduction methods have become instrumental in understanding how network structure and interactions shape the collective dynamics and the emergence  of synchrony. We review this progress based on concrete examples and outline possible limitations. Finally, we discuss how linking the reduced models with experimental data can guide the way towards the development of new treatment approaches, for example, for neurological disease.




\section{Introduction}

Many systems in neuroscience and biology are governed on different levels by interacting periodic processes~\cite{Winfree2001}. Networks of coupled oscillators provide models for such systems. Each node in the network is an oscillator (a dynamical process) and the network structure encodes which oscillators interact with each other~\cite{Strogatz2001}. In neuroscience, individual oscillators could be single neurons in microcircuits or neural masses on a more macroscopic level~\cite{Breakspear2017}. Other prominent examples in biology include cells in heart tissue~\cite{Liu1997}, flashing fireflies~\cite{Buck1968}, the dynamics of cilia and flagella~\cite{Gilpin2020}, gait patterns of animals~\cite{Collins1993} or humans~\cite{Strogatz2005}, cells in the suprachiasmatic nucleus in the brain generating the master clock for the circadian rhythm~\cite{Strogatz1987,Leloup2008,Smolen2009}, 
hormone rhythms~\cite{Zavala2019},
suspensions of yeast cells undergoing metabolic oscillations~\cite{Ghosh1971,Dano1999}, and life cycles of phytoplankton in chemostats~\cite{Massie2010}. 

The functionality---whether function or dysfunction---of these networks depends on the \emph{collective dynamics} of the interacting oscillatory nodes.
Hence, one major challenge is to understand how the underlying network shapes these collective dynamics.
In particular, one would like to understand how the interplay of network properties (for example, coupling topology and the strength of interactions) and characteristics of the individual nodes shape the emergent dynamics.
The question of relating network structure and dynamics is particularly pertinent in the study of large-scale brain dynamics: For example, one can investigate how emergent functional connectivity (a dynamical property) arises from specific structural connectomes~\cite{Honey2007,Honey2010}, and how each of these relates to cognition or disease.  
Progress in this direction not only aims to identify how healthy or pathological dynamics is linked to the network structure, but also to develop new treatment approaches~\cite{Fornito2015,Bassett2017,Kuhlmann2018,Goodfellow2016}. 

One of the most prominent collective behaviors of an oscillator network occurs when nodes synchronize and oscillate in unison~\cite{Strogatz2004sync,Glass2001,Dorfler2014}; indeed, most of the examples given above display synchrony in some form which appears to be essential to the proper functioning of biological life processes.
Here we think of synchrony in a general way: It can come in many varieties, including \emph{phase synchrony} where the state of different oscillators align exactly, or \emph{frequency synchrony} where the oscillators' frequencies coincide. Synchrony may be \emph{global} across the entire network or \emph{localized} in a particular part---the rest of the network being nonsynchronized---thus giving rise to \emph{synchrony patterns}.
How exactly the dynamics of synchrony patterns in an oscillator network relate to its functional properties is still not fully understood. In the brain, there are a wide range of rhythms but the presence of dominant rhythms in different frequency bands indicate that some level of synchrony is common at multiple scales~\cite{Kahana2006,Lehnertz2017}. Indeed, synchrony has been associated with solving functional tasks including, but not limited to, memory~\cite{Fell2011}, computational functions~\cite{Fries2009}, cognition~\cite{Wang2010}, attention~\cite{Singer1995,Fries2005}, routing of information~\cite{Fries2005,Kirst2016,DeschleMartens2019}, control of gait and motion~\cite{Marder2001}, or breathing~\cite{Smith1991,Butera1999}. As a specific example, coordination of dynamics at the theta frequency (4--12Hz) between hippocampus and frontal cortex is enhanced in spatial memory tasks~\cite{Jones2005}.
At the same time, abnormal synchrony patterns are associated with malfunction in disorders such as epilepsy and Parkinson's disease~\cite{Uhlhaas2006,Hammond2007,Lehnertz2009,Rummel2012}; evolving patterns of synchrony can for example be observed in electroencephapholographic (EEG) recordings throughout epileptogenesis in mice~\cite{Sowinski2019}.

Using a detailed model of each node and a large number of nodes in the network, relating network structure and dynamics is a daunting task. Hence, simplifying analytical reduction methods are needed that---rather than being purely computational---yield a mechanistic understanding of the inherent processes leading to a certain dynamic macroscopic behavior. If many biologically relevant state variables are considered in a microscopic model, each node is represented by a high-dimensional dynamical system by itself. Hence, a common approach is to simplify the description of each oscillatory node to its  simplest form, a \emph{phase oscillator}; in the reduced system the state of each oscillator is given by a single periodic phase variable that captures the state of the periodic process. In this case, biologically relevant details are captured by the evolution of the phase variable and its interaction with the phases of the other nodes. There are two important ways to get to a phase description of an oscillator network, both of which are common tools used, for example, in computational neuroscience; see~\cite{Ashwin2016,Pietras2019} for recent reviews. First, under the assumption of weak coupling one can go through a process of phase reduction to obtain a phase description~\cite{Ashwin1992,Hansel1993a,Hoppensteadt1997,Brown2004,Nakao2015,Monga2018}. Second, one can---based on the biophysical properties of the system---impose a phase model such as the Kuramoto model~\cite{Cabral2011} or a network of Theta neurons~\cite{luke2014}.

The main topic of this paper is an introduction to and a review of recent advances in exact mean-field reductions for networks of coupled oscillators. The main achievement is that for certain classes of oscillator networks, it is possible to replace a large number of nodes by a \emph{collective} or \emph{mean-field variable} that describe the network evolution exactly---thereby reducing the complexity of the problem immensely. In the neuroscientific context, each subpopulation may represent different populations of neurons that may exhibit temporal patterns of synchronization or activity~\cite{Honey2007,Honey2010,Britz2010}. Of course, mean-field approaches motivated by statistical physics have a long history in computational neuroscience to approximate the dynamics of large ensembles of units; see, for example,~\cite{Destexhe2009,Gupta2018} and references therein. They have been useful to elucidate, for example, dynamical mechanisms behind the emergence of rhythms in the gamma frequency band, such as the emergence of pyramidal-interneuronal gamma (PING) rhythm~\cite{borgers2003} or the interplay between different brain areas (for example, through phase-phase, phase-amplitude and amplitude comodulation) that can lead to frequency entrainment~\cite{Buzsaki2012a}. In terms of classical mean-field approaches, the pioneering works by Wilson and Cowan~\cite{wilson1973} and Amari~\cite{amari1977} stand out: They derived heuristic equations for average neural population dynamics that are still widely used in neural modeling. Specifically, such models disregard fluctuations of individual units\footnote{naturally, any mean-field approach (including the Ott--Antonsen reduction we consider here) that disregards (finite-size) fluctuations to analyze \emph{finite} networks cannot capture dynamical effects where these fluctuations are a characteristic property. This includes the balanced state~\cite{vanVreeswijk1996,vanVreeswijk1998}.} and arrive at equations that approximate the evolution of means.
By contrast, the exact mean-field reductions we discuss here, the Ott--Antonsen reduction and the Watanabe--Strogatz reduction, can be employed not only for infinite networks also for networks of finitely many oscillators.
While these reductions only apply to specific classes of systems---and from a mathematical perspective reflect the special structure of these systems---they include models that have been widely used in neuroscience and beyond, such as the Kuramoto model.
Compared to heuristic mean-field approximations, the resulting reduced equations are exactly derived from the microscopic equations of individual oscillators and thus capture properties of individual oscillators; because of this property these reduced equations have been referred to as being \emph{next-generation models}~\cite{Coombes2016}. Employing these models in modeling tasks provides a powerful opportunity to bridge the dynamics on microscopic and macroscopic scales.

To illustrate the mean-field reductions and their applicability, we focus here on networks that are organized into distinct (sub)populations because of their practical importance\footnote{Such networks may be thought of as ``networks of networks''~\cite{Barreto2008,Kivela2014}.}. The mean-field reductions allow one to replace each subnetwork by a (low-dimensional) set of collective variables to obtain a set of dynamical equations for these variables. This set of mean-field equations describes the system exactly. For the classical Kuramoto model, which is widely used to understand synchronization phenomena, we will see below that the collective state is captured by a two-dimensional mean-field variable that encodes the synchrony of the population. Reducing to a set of mean-field equations provides a simplified---but often still sufficiently complex---description of the network dynamics that can be analyzed by using dynamical systems techniques~\cite{Strogatz1994book,Izhikevich2007}. We will outline the classes of models for which the mean-field reductions apply and illustrate how these reduction techniques have been instrumental in the last decade to illuminate how the network properties relate to dynamical phenomena. We give a number of concrete examples, from Kuramoto's problem about the emergence of synchrony in oscillator populations to the emergence of PING rhythms based on microscopic properties of neuronal networks.

There are many important questions and aspects that we cannot touch upon in this review, and we refer to already existing reviews and literature instead. 
First, we only consider oscillator networks where each (microscopic) node has a first-order description by a single phase variable. We will not cover other microscopic models such as second-order phase oscillators or oscillators with a phase and amplitude\footnote{Note, however, that the mesoscopic description in terms of collective variables of each subnetwork can have a ``phase'' and ``amplitude'' such as the mean phase and the amount of synchrony.} which can give rise to richer dynamics.
Second, we do not comment on the validity of a phase reduction; for more information see for example~\cite{Hoppensteadt1997,Monga2018}. 
Third, the reductions we discuss have been essential to understand the emergence of synchrony patterns where coherence and incoherence coexist, also known as ``chimeras.'' Here, we only mention results relevant to the dynamics of coupled oscillator populations and refer to~\cite{PanaggioAbramsReview2015,Scholl2016,Omelchenko2018} for recent reviews on chimeras.
Fourth, the results mentioned here relate to results from network science~\cite{Porter2014,Rodrigues2016}. In particular, properties of the graph underlying the dynamical network relate to synchronization dynamics~\cite{Pecora1998,Barahona2002,Pereira2014}. Moreover, we also typically assume that the network structure is static and does not evolve over time. However, time-dependent network structures are clearly of interest---in particular in the context of plastic neural connectivity and neurological disease. An approach to these issues from the perspective of network science are temporal networks~\cite{Holme2012} while asynchronous networks take a more dynamical point of view; see~\cite{Bick2015} and references therein. 
Fifth, we restrict ourselves to deterministic dynamics where noise is absent. From a mathematical point of view, noise can simplify the analysis and recent results show that similar reduction methods apply~\cite{Tyulkina18,Goldobin2018,Goldobin2018b}.
Finally, it is worth noting that other reduction approaches for oscillator networks have recently been explored~\cite{Gottwald2015,Skardal2017,Hannay2018}.

This review is written with a diverse readership in mind, ranging from mathematicians to computational biologists who want to use the reduced equations for modeling. 
In fact, this review is intended to have aspects of a tutorial and to provide an introduction to the Ott--Antonsen and Watanabe--Strogatz reductions as exact mean-field reductions and outlining what type of questions they have been instrumental in answering: We include three explicit examples how these mean-field reductions can be helpful in giving insights into the collective dynamics of (neuronal) oscillator networks.

In the following, we provide an outline how to approach this paper.
The next section sets the stage by introducing the notion of a sinusoidally coupled network and we summarize the main oscillator models we relate to throughout the paper; these include the Kuramoto model and networks of Theta neurons (which are equivalent to Quadratic Integrate and Fire (QIF) neurons).
In the third section, we give a general theory for the mean-field reductions and discuss their limitations: The methods include the Ott--Antonsen reduction for the mean-field limit of nonidentical oscillators and the Watanabe--Strogatz reduction for finite or infinite networks of identical oscillators. This section includes a certain level of mathematical detail to understand the ideas behind the derivation of the reduced equations (mathematically dense sections are marked with the symbol ``*'' and may be skipped at first reading). If you are mainly interested in applying the reduced equations, you may want to skip ahead to Sections~\ref{sec:OACommonlyUsed} and~\ref{sec:WSCommonlyUsed}, which summarize the reduced equations for the models we study throughout the paper.
In the fourth section, we apply the reductions and emphasize how they are useful to understand how synchrony and patterns of synchrony emerge in such oscillator networks. This includes a number of concrete examples.
Since most of these considerations are theoretical and computational, we discuss in the last section how the mean-field reductions can be used to solve neuroscientific problems and be linked with experimental data.
We conclude with some remarks and highlighting a number of open problems.

\subsection{List of symbols}
The following symbols will be used throughout this paper.

\begin{longtable}{ll}
$\N$ & The positive integers\\
$\Tor$ & The circle of all phases $\R/2\pi\Z$ (or $[0, 2\pi]$ with $0\equiv2\pi$)\\
$\C$ & The complex numbers\\
$i$ & Imaginary unit $\sqrt{-1}$\\
$\Re{w}, \Im{w}$ & Real part and imaginary part of a complex number~$w\in\C$\\
$\bar w$ & Complex conjugate of~$w\in\C$\\
$\maxpop$ & Number of oscillator populations in the network\\
$\sigma,\tau$ & Population indices in $\sset{1,\dotsc,\maxpop}$\\
$\maxdim$ & Number of oscillators in each population\\
$k,j,l,\dotsc$ & Oscillator indices in $\sset{1,\dotsc,\maxdim}$\\
$\theta_{\sigma,k}$ & Phase of oscillator~$k$ in population~$\sigma$\\
$\kappa, \kgj, \kg$ & Coupling strength between neural oscillators\\
$Z_\sigma$ & Kuramoto order parameter of population~$\sigma$\\
$R_\sigma$ & The level of synchrony~$\abs{Z_\sigma}$ of population~$\sigma$\\
$z_\sigma,\Psi_\sigma$ & Bunch variables of population~$\sigma$\\
$\dot x$ & The time derivative $\frac{\ud x}{\ud t}$ of~$x$\\
\end{longtable}

\section{Sinusoidally coupled phase oscillator networks}
\label{sec:OscillatorModels}

The state of each node in a phase oscillator network is given by a single phase variable. Such networks may be obtained through a phase reduction or may be abstract models in their own right as in the case of the Theta neuron below. Consider a population~$\sigma$ of~$\maxdim$ oscillators where the state of oscillator~$k$ is given by a phase $\theta_{\sigma,k}\in\Tor:=\R/2\pi\Z$; if there is only a single population, we drop the index~$\sigma$.
Without input, the phase of each oscillator~$(\sigma, k)$ advances at its \emph{intrinsic frequency}~$\omega_{\sigma, k}\in\R$.
Input to oscillator~$(\sigma, k)$ is determined by a field~$H_{\sigma,k}(t)\in\C$ and modulated by a sinusoidal function; 
this field could be an external drive or network interactions between oscillators both within population~$\sigma$ or other populations~$\tau$.
Specifically, we consider oscillator networks whose phases evolve according to
\begin{equation}\label{eq:Microscopic}
  \dot\theta_{\sigma,k} = \omega_{\sigma, k} + \Im{H_{\sigma,k}(t) e^{-i\theta_{\sigma,k}}}.
\end{equation}
Since the effect of the field on oscillator~$(\sigma, k)$ is mediated by a function with exactly one harmonic, $e^{-i\theta_{\sigma,k}}$, we call the oscillator populations \emph{sinusoidally coupled}\footnote{In equation~\eqref{eq:Microscopic} the coupling is through a pure first harmonic. However, the results presented here are generally valid for coupling through any pure single harmonic of higher order; see for example~\cite{Gong2019,Skardal2019a}.}. 

While we allow the intrinsic frequency and the driving field to depend on the oscillator to a certain extent (i.e., oscillators are nonidentical), we will henceforth assume that all oscillators within any given population~$\sigma$ otherwise are \emph{(statistically) indistinguishable}, i.e., the properties of each oscillator in a given population are determined by the same distribution. Specifically, suppose that the properties of each oscillator are determined by a certain parameter~$\eta_{\sigma,k}$. This is for example the case for the Theta neurons described further below, each of which has an individual level of excitability as a parameter. Let us formulate this more precisely. Suppose that we let both the intrinsic frequencies and the field be functions of this parameter, i.e., $\omega_{\sigma, k} = \omega_\sigma(\eta_{\sigma,k})$, $H_{\sigma,k}(t) = H_\sigma(t; \eta_{\sigma,k})$. The oscillators of a given population are then indistinguishable if, for a given population~$\sigma$, all~$\eta_{\sigma,k}$ are random variables sampled from a single probability distribution with density~$h_\sigma(\eta)$. In the special case that~$\eta_{\sigma,k}=\eta_{\sigma,j}$ for $j\neq k$ (in this case~$h_\sigma$ is a delta-distribution) we say that the oscillators are \emph{identical}.

Phase oscillator networks of the form~\eqref{eq:Microscopic} include a range of well-known (and well-studied) models. These range from particular cases of Winfree's model~\cite{winfree1967} to neuron models. In the following we discuss some important examples that we will revisit in more detail throughout this paper.

\subsection{The Kuramoto model}
\label{sec:KuramotoModel}

Kuramoto originally studied synchronization in a network of~$\maxdim$ globally coupled nonidentical (but indistinguishable) phase oscillators~\cite{Kuramoto1984}; see~\cite{Strogatz2000} for an excellent survey of the problem and its historical background. 
Kuramoto originally investigated the \emph{onset of synchronization} in a network composed of only a single population of oscillators indexed by $k\in\sset{1, \dotsc, \maxdim}$ with phases~$\theta_k$ (here we drop the population index~$\sigma$).
The oscillator phases evolve according to
\begin{equation}\label{eq:Kuramoto}
\dot\theta_k = \omega_k+\frac{K}{\maxdim}\sum_{j=1}^\maxdim\sin(\theta_j-\theta_k)
\end{equation}
with distinct intrinsic frequencies~$\omega_k$ that are sampled from some unimodal frequency distribution.
Here the parameter~$K$ is the coupling strength between oscillators and the coupling is mediated by the sine of the phase difference between oscillators. If coupling is absent ($K=0$), each oscillator advances with its intrinsic frequency~$\omega_k$. 

The macroscopic state of the population is characterized by the complex-valued \emph{Kuramoto order parameter}\footnote{The order ``parameter''~$Z$ is an observable which encodes the state of the system, and should not be confused with a system parameter.}
\begin{equation}\label{eq:KuramotoOPFin}
\OP = Re^{i\phi} =\frac{1}{\maxdim}\sum_{j=1}^{\maxdim}e^{i\theta_{j}},
\end{equation}%
representing the mean of all phases on the unit circle.
\begin{figure}
\begin{center}
\includegraphics[width=0.8\textwidth]{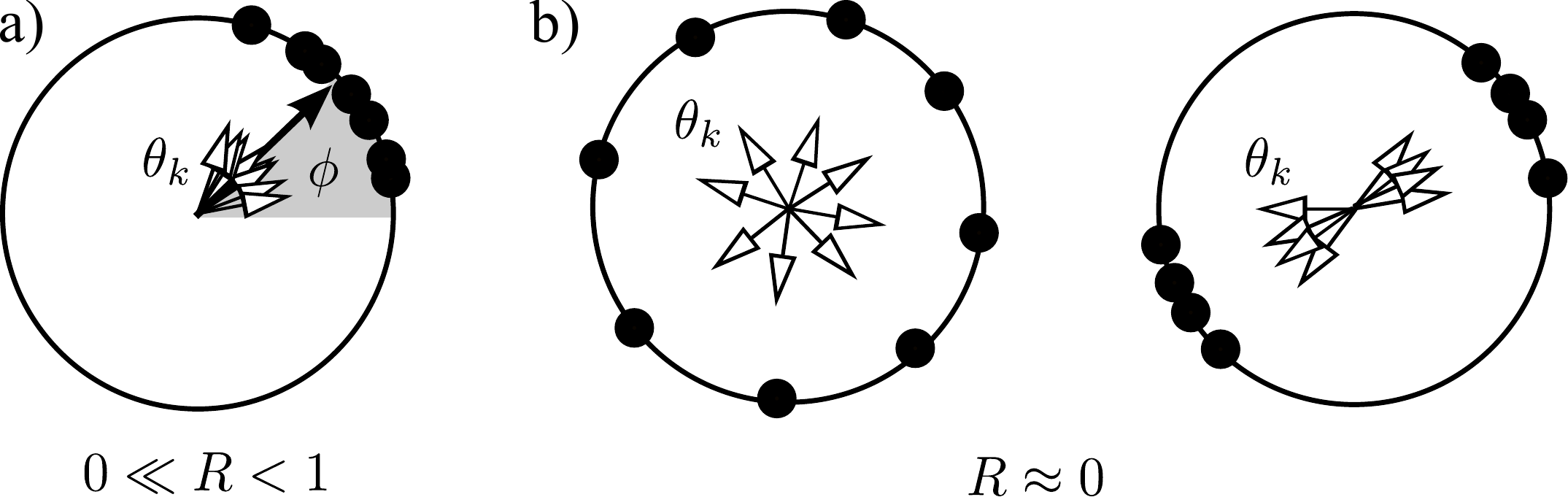}
\smallskip
\end{center}
\caption{\label{fig:OP}The Kuramoto order parameter~\eqref{eq:KuramotoOPFin} encodes the level of synchrony of a phase oscillator population. The state of each oscillator is given by a phase~$\theta_k$ (black dot, empty arrow) on the circle~$\Tor$. Panel~(a) shows a configuration with high synchrony where $R=\abs{\OP}\lessapprox 1$. Panel~(b) shows two configurations with $R=\abs{\OP}\gtrapprox 0$: one where the oscillators are approximately uniformly distributed on the circle, the other one where they are organized into two clusters.}
\end{figure}%
Its magnitude~$R=\abs{\OP}$ describes the level of synchronization of the oscillator population, see Fig.~\ref{fig:OP}: On the one hand, $R=1$ if and only if all oscillators are \emph{phase synchronized}, that is, $\theta_k=\theta_j$ for all $k$ and $j$; on the other hand, we have $R=0$ if, for example, the oscillators are evenly distributed around the circle. 
The argument~$\phi$ of the Kuramoto order parameter~$Z$ (which is well-defined for $Z\neq 0$) describes the ``average phase'' of all oscillators, that is, it describes the average position of the oscillator crowd on the circle of phases.

Kuramoto observed the following macroscopic behavior: For~$K$ small, the system converges to an incoherent stationary state with $R\approx 0$. As~$K$ is increased past a critical coupling strength~$K_c$, the system settles down to a state with partial synchrony, $R>0$. As the coupling strength is further increased, $K\to\infty$, oscillators become more and more synchronized, $R\to 1$.

The Kuramoto model~\eqref{eq:Kuramoto} is an example of a sinusoidally coupled phase oscillator network. Using Euler's identity $e^{i\phi}=\cos(\phi)+i\sin(\phi)$, we have
\[\dot\theta_k = \omega_k+\Im{\frac{K}{\maxdim}\sum_{j=1}^\maxdim e^{i(\theta_j-\theta_k)}}
=\omega_k+\Im{K\OP(t) e^{-i\theta_k}},
\]
where the Kuramoto order parameter~$\OP(t) = \OP(\theta_1(t), \dotsc, \theta_\maxdim(t))$, as defined in~\eqref{eq:KuramotoOPFin}, depends on time through the phases. Hence, the Kuramoto model~\eqref{eq:Kuramoto} is equivalent to~\eqref{eq:Microscopic} with $H(t)=K\OP(t)$ and the interactions between oscillators are solely determined by the Kuramoto order parameter~$\OP(t)$. Such a form of network interaction is also called \emph{mean-field coupling} since the drive~$H(t)$ to a single oscillator is proportional to a mean field, that is, the average formed from the states of all oscillators in the network.

\medskip
\noindent\textbf{Problem 1:\ }\label{problem1}%
How can mean-field reductions elucidate Kuramoto's original problem of the onset of synchronization in an infinitely large population of oscillators?
We will revisit this problem in \hyperref[example1]{\emph{Example~1}} below.

\subsection{Populations of Kuramoto--Sakaguchi oscillators}
\label{sec:KuramSakaguchi}

Sakaguchi generalized Kuramoto's model by introducing an additional  phase-lag (or phase-frustration) parameter which approximates a time delay in the interactions between oscillators~\cite{Sakaguchi1986,PanaggioAbramsReview2015}. While Sakaguchi originally considered a single population of oscillators, here we generalize to multiple interacting populations. Specifically, we consider the dynamics of~$\maxpop$ populations of~$\maxdim$ Kuramoto--Sakaguchi oscillators, where the phase of oscillator~$k$ in population~$\sigma$ evolves according to
\begin{equation}\label{eq:PopKuramotoSakaguchi}
\dot\theta_{\sigma,k} = \omega_{\sigma,k}+\sum_{\tau=1}^\maxpop\frac{K_{\sigma\tau}}{\maxdim}\sum_{j=1}^\maxdim \sin(\theta_{\tau,j}-\theta_{\sigma,k}-\alpha_{\sigma\tau}),
\end{equation}
and where $K_{\sigma\tau}\geq 0$ is the \emph{coupling strength} and $\alpha_{\sigma\tau}$ is the \emph{phase lag} between populations~$\sigma$ and~$\tau$.%
\footnote{If only one population is considered, $\maxpop=1$, we simply write $\alpha_{\sigma\tau}=\alpha$ and  $K_{\sigma\tau}=K$; this corresponds to the Kuramoto--Sakaguchi model. Furthermore, if $\alpha=0$, we recover the Kuramoto model~\eqref{eq:Kuramoto}. Here, we regard the number of populations~$\maxpop$ to be fixed; to take a limit~$\maxpop\to\infty$ one should assume that the coupling strengths~$K_{\sigma\tau}$ scale appropriately.}
The function~$g_{\sigma\tau}(\phi)=K_{\sigma\tau}\sin(\phi-\alpha_{\sigma\tau})$ mediates the interactions between oscillators, and we refer to it as the \emph{coupling function}; later on we will also briefly touch upon what happens if the sine function is replaced by a  \emph{general} periodic coupling function.
As in the Kuramoto model, an important point is that the influence between oscillators $(\tau,j)$ and $(\sigma,k)$ depends only on their phase \emph{difference} (rather than explicitly on their phases)\footnote{Such form of interactions typically arise in an additional averaging step performed after the phase reduction~\cite{Swift1992,Ashwin2016}.}. Thus, this form of interaction only depends on the relative phase between oscillator pairs rather than the absolute phases. An important consequence is that the dynamics of equations~\eqref{eq:PopKuramotoSakaguchi} do not change if we consider all phases in a different reference frame. For example, going into a reference frame rotating at constant frequency $\omf\in\R$ corresponds to the transformation $\theta_{\sigma,k}\mapsto \theta_{\sigma,k}-\omf\,t$, only shifts all intrinsic frequencies by~$\omf$ rather than changing the dynamics qualitatively\footnote{One may also consider co-rotating frames with time-dependent frequencies. For example, for any given oscillator~$(\tau, j)$ one can choose a co-rotating frame in which its phase $\theta_{\tau, j}$ appears stationary via the transform $\theta_{\sigma,k}\mapsto \theta_{\sigma,k}-\theta_{\tau,j}$. This transformation changes the structure of~\eqref{eq:PopKuramotoSakaguchi} but it does not affect the qualitative dynamics.}.

The network~\eqref{eq:PopKuramotoSakaguchi} of~$\maxpop$ interacting populations of Kuramoto--Sakaguchi oscillators is a sinusoidally coupled oscillator network. The amount of synchrony in population~$\sigma$ is then determined by the Kuramoto order parameter~\eqref{eq:KuramotoOPFin} for population~$\sigma$,
\begin{equation}\label{eq:KuramotoOPpopulation}
\OP_\sigma = \frac{1}{\maxdim}\sum_{j=1}^{\maxdim}e^{i\theta_{\sigma,j}}.
\end{equation}
Combining coupling strength and phase lag, we define the complex interaction parameter $c_{\sigma\tau} := K_{\sigma\tau} e^{-i\alpha_{\sigma\tau}}$ between populations~$\sigma$ and~$\tau$.
By the same calculation as above, the network~\eqref{eq:PopKuramotoSakaguchi} is equivalent to~\eqref{eq:Microscopic} with constant intrinsic frequencies~$\omega_{\sigma,k}$ and driving field
\begin{align}\label{eq:DriveKS}
H_\sigma = \sum_{\tau=1}^\maxpop c_{\sigma\tau}\OP_\tau,
\end{align}
being a linear combination of the mean fields of the other populations.

Networks of Kuramoto--Sakaguchi oscillators have been used as models for synchronization phenomena. In neuroscience, individual oscillators can represent neurons~\cite{cumin2007} or large numbers of neurons in neural masses~\cite{breakspearH2010,Cabral2011,Schmidt2014}. In the framework of the model~\eqref{eq:PopKuramotoSakaguchi}, the populations can be thought of as~$\maxpop$ neural masses.
In contrast to models where neural masses only have a phase, here, the macroscopic state of each population (neural mass) is determined by an amplitude (the level of synchrony $R_\sigma := |Z_\sigma|$) and an angle (the average phase $\phi_\sigma := \arg{Z_\sigma}$).

\subsection{Theta and quadratic integrate and fire neurons}

\begin{figure}
\begin{center}
\includegraphics[width=0.8\textwidth]{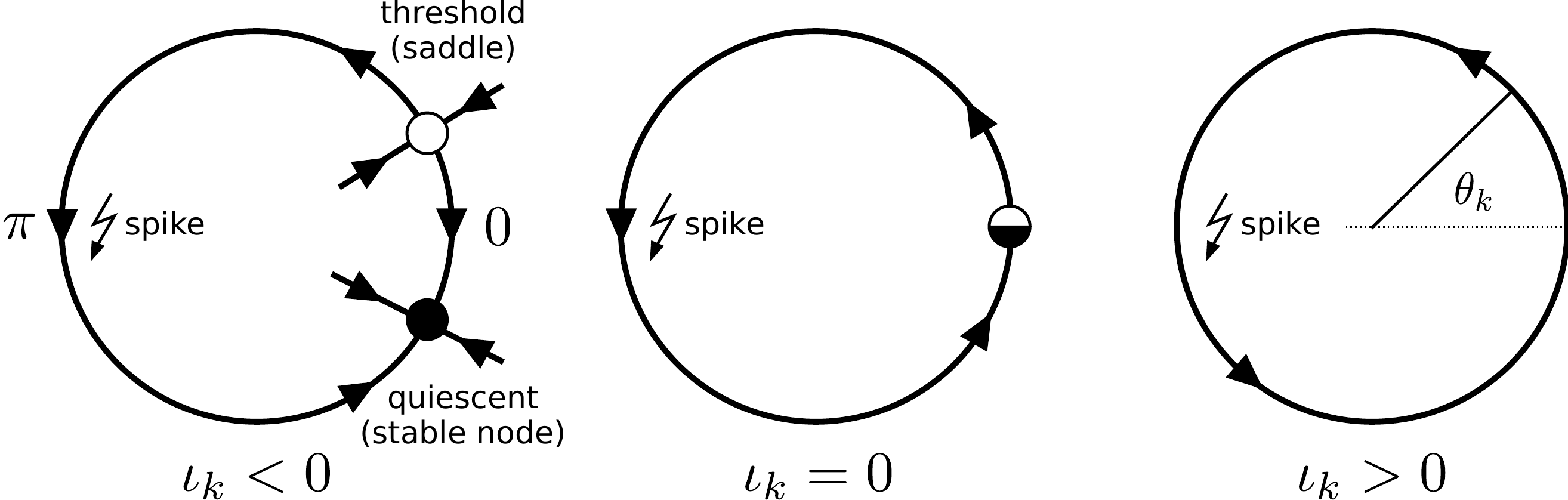}
\smallskip
\end{center}
\caption{\label{fig:ThetaNeuronSNIC}
A Theta neuron~\eqref{eq:dthdtD} with phase~$\theta_k$ subject to constant input~$I$ undergoes a saddle node bifurcation on an invariant circle (SNIC) as the quantity~$\iota_k=\eta_k+\kappa I$ is varied. The neuron spikes if its phase~$\theta_k$ crosses $\theta_k=\pi$. If $\iota_k<0$ the Theta neuron is excitable: the phase will relax to the stable equilibrium and a phase perturbation of the phase across the saddle equilibrium (its threshold) will lead to a single spike before returning to equilibrium. For $\iota_k>0$, the Theta neuron is spiking periodically.}
\end{figure}

\paragraph{Theta neurons.} The Theta neuron is the normal form of the saddle-node-on-invariant-circle (SNIC) or saddle-node-infinite-period (SNIPER) bifurcation~\cite{Ermentrout2008} as shown in Fig.~\ref{fig:ThetaNeuronSNIC}: At the excitation threshold, a saddle and a node coalesce on an invariant circle (i.e., limit cycle of the neuron). Its state is described by the phase~$\theta\in\Tor$ on the invariant circle and we use the convention\footnote{This convention is in line with the firing of the equivalent quadratic integrate and fire neuron introduced below.} that the neuron fires (it emits a spike) when the phase crosses $\theta=\pi$ (Fig.~\ref{fig:ThetaNeuronSNIC}). The Theta neuron is a valid description of the dynamics of any neuron model undergoing this bifurcation, in some parameter neighborhood of the bifurcation. The Theta neuron is also a canonical Type~1 neuron~\cite{ermentrout1986parabolic}.

Consider a single population of Theta neurons (hence we drop the population index~$\sigma$) whose phases evolve according to
\be
   \dot\theta_k=1-\cos{\theta_k}+(1+\cos{\theta_k})(\eta_k+\kappa I),\label{eq:dthdtD}
\ee
where~$\eta_k$ is the excitability of neuron~$k$ sampled from a probability distribution,~$\kappa$ is the coupling strength, and $I$~is an \emph{input current}---this could result from external input (driving) or network interactions.
A population of Theta neurons~\eqref{eq:dthdtD} is a sinusoidally coupled system 
of the form~\eqref{eq:Microscopic} with
\begin{align}\label{eq:DriveTheta}
\omega_k &= 1+\eta_k+\kappa I, & H_k &= i(\eta_k+\kappa I-1).
\end{align}
The dependence of $H, \omega$ on the excitability parameters~$\eta_k$ can be made explicit by writing $\omega_k = \omega(\eta_k)$, $H(t) = H(t; \eta_k)$. Thus, results for models of the form~\eqref{eq:Microscopic} will also apply to networks of Theta neurons.

The Theta neuron was introduced in 1986~\cite{ermentrout1986parabolic} and has since then been widely used in neuroscience. We refer for example to~\cite{Ermentrout2010,Gerstner2014} for a general introduction and only list a few concrete applications here. For example, Monteforte and Wolf~\cite{Monteforte2010} used these neurons as canonical type I neuronal oscillators in their study of chaotic dynamics in large, sparse balanced networks. The papers~\cite{osan2001,ermentrout2002} considered spatially extended networks of Theta neurons and the authors were specifically interested in traveling waves of activity in these networks. More recently, other authors have used some of the techniques for dimensional reduction reviewed in the present paper to study infinite networks of Theta neurons~\cite{Luke2013,Laing2014}. We will discuss these reduction methods in detail further below.

\medskip
\noindent\textbf{Problem 2:\ \label{problem2}}%
What different dynamics are possible in a single population of globally coupled Theta neurons with pulsatile coupling? What is the onset for firing of neurons? We will revisit this problem in \hyperref[example2]{\emph{Example~2}} below.

\paragraph{Quadratic Integrate and Fire neurons.}
The Theta neuron model is closely related to the Quadratic Integrate and Fire (QIF) neuron model~\cite{gutkin2015} whose state is given by a membrane voltage $V\in(-\infty, +\infty)$. More precisely, using the transformation $V_k=\tan{(\theta_k/2)}$ the population of Theta neurons~\eqref{eq:dthdtD} becomes a population of QIF neurons, where the membrane voltage~$V_k$ of neuron~$k$ evolves according to
\be \label{eq:QIF}
   \dot V_k=V_k^2+\eta_k+\kappa I.
\ee
Here we use the rule that the neuron fires (it emits a spike) if its voltage reaches $V_k(t^-)=+\infty$ and then the neuron is reset to $V_k(t^+)=-\infty$.

QIF neurons have been widely used in neuroscientific modeling; see~\cite{Ermentrout2010,Gerstner2014} for a general introduction and~\cite{latham2000,hansel2001,brunel2003,kopell2004} for a few examples in the literature where QIF neurons are employed. They have the simplicity of the more common leaky integrate-and-fire model in the sense of having only one state variable (the voltage), but are more realistic in the sense of actually producing spikes in the voltage trace~$V(t)$.

\medskip
\noindent\textbf{Problem 3:\ }\label{problem3}%
How does a network of neurons respond to a transient stimulus? Specifically, if this neuronal network is modeled by a heterogeneous network of all-to-all coupled QIF neurons. This is a pertinent question, for example, if stimulation is used for therapeutic purposes such as in Deep Brain Stimulation. 
We will revisit this problem in \hyperref[example3]{\emph{Example~3}} below.

\section{Exact mean-field descriptions for sinusoidally coupled phase oscillators}
\label{sec:CollectiveVars}

In this section, we review how sinusoidally coupled phase oscillator networks~\eqref{eq:Microscopic} can be simplified using mean-field reductions. Under specific assumptions (detailed further below) we derive low-dimensional system of ordinary differential equations for macroscopic mean-field variables that describe the evolution of sinusoidally coupled phase oscillator networks~\eqref{eq:Microscopic} exactly. This is in contrast to reductions that are only approximate or only valid over short time scales. Thus, these reduction methods facilitate the analysis of the network dynamics: rather than looking at a complex, high-dimensional network dynamical system (or its infinite-dimensional mean-field limit) we can analyze simpler, low-dimensional equations. For example, for the infinite-dimensional limit of the Kuramoto model, we obtain a closed system for the evolution of~$\OP$, a two-dimensional system (since~$Z$ is complex). While the Kuramoto model is particularly simple, the methods apply for general driving fields~$H_{\sigma,k}$ that could contain delays or depend explicitly on time. We give concrete examples in Section~\ref{sec:UsingMFR} below, where we apply the reduction techniques.

Importantly, these mean-field reductions also apply to oscillator networks which are equivalent to~\eqref{eq:Microscopic}. In particular, this applies to neural oscillators: The QIF neuron and the Theta neuron are equivalent as discussed above. Consequently, rather than assuming a model for a neural population (e.g.,~\cite{Cabral2011}), we actually obtain an exact description of interacting neural populations in terms of their macroscopic (mean-field) variables.

\subsection{Ott--Antonsen reduction for the mean-field limit of nonidentical oscillators}
\label{sec:RedOA}

The Ott--Antonsen reduction applies to the mean-field limit of populations of indistinguishable sinusoidally coupled phase oscillators~\eqref{eq:Microscopic}. First, we first outline the basic steps to derive the equations and highlight the assumptions made along the way; this section contains mathematical details and may be omitted on first reading. We then summarize the Ott--Antonsen equations for the models described in the previous section.

\subsubsection{*Derivation of the reduced equations}
\label{sec:RedOAPhaseOsc}

Consider the dynamics of the (mean-field) limit of~\eqref{eq:Microscopic} with infinitely many oscillators, $\maxdim\to\infty$. Note that while the population index~$\sigma$ is seen as discrete in this paper, it is also possible to apply the reduction to continuous topologies of populations such as rings; cf.~\cite{Laing2009,Omel'chenko2013}. To simplify the exposition, we consider the classical case where the intrinsic frequency is the random parameter, $\omega_{\sigma,k}=\eta_{\sigma, k}$, and that the driving field is the same for all oscillators in any population, $H_{\sigma,k} = H_{\sigma}$; for details on systems with explicit parameter dependence (such as Theta neurons) see~\cite{Montbrio2015,Pietras2016a}. Hence, suppose that the intrinsic frequencies~$\omega_{\sigma,k}$ are randomly drawn from a distribution with density~$h_\sigma(\omega)$ on~$\R$. In the mean-field limit, the state of each population at time~$t$ is not given by a collection of oscillator phases, but rather by a probability density~$f_\sigma(\omega, \vth; t)$ for an oscillator with intrinsic frequency~$\omega\in\R$ to have phase~$\vth\in\Tor$; see~\cite{Mardia1999} for general properties of such distributions and statistics on the circle. For a set of phases $B\subset\Tor$ the marginal $\int_B\int_\R f_\sigma(\omega, \vth; t)\udi\omega\udi\vth$ determines the fraction of oscillators whose phase is in~$B$ at time~$t$. Moreover, we have $\int_\Tor f_\sigma(\omega, \vth; t)\udi\vth = h_\sigma(\omega)$ for all times~$t$
by our assumption that the intrinsic frequencies do not change over time.

Conservation of oscillators implies that the dynamics of the mean-field limit of~\eqref{eq:Microscopic} is given by the transport equation\footnote{If the oscillators are subject to noise, the continuity equation is a Fokker--Planck equation which contains an additional diffusive term~\cite{Sakaguchi1988,Crawford1994,Acebron2005,Lai2013a}.}
\begin{align}\label{eq:TransportEq}
\frac{\partial f_{\sigma}}{\partial t} + \frac{\partial}{\partial\vth}\left(v_\sigma f_\sigma\right)&=0 &\text{with}&& v_\sigma=\omega_\sigma + \Im{H_\sigma(t)e^{-i\vth}}.
\end{align}
Because oscillators are conserved\footnote{Transport equations are common in physics. There they are also known as the continuity equation (or Liouville equation in classical statistical physics describing the ensemble evolution in time) and play the important role of describing conservation laws. To visualize, in the context of fluid dynamics, the density in~\eqref{eq:TransportEq} plays the role of a mass density and \eqref{eq:TransportEq} then implies that the total mass in the system is a conserved quantity~\cite{Landau1959}.}, the change of the phase distribution over time is determined by the change of phases given by the velocity~$v_\sigma$ through~\eqref{eq:Microscopic} at time~$t$ of an oscillator with phase~$\vth$ and intrinsic frequency~$\omega$. While the transport equation for the mean-field limit originally appears in Refs.~\cite{Sakaguchi1988,Strogatz1991}, it can be rigorously derived from a measure-theoretic perspective as a Vlasov limit~\cite{Lancellotti2005}.

Before we discuss how to find solutions for the transport equation~\eqref{eq:TransportEq}, it is worth noting that it has been analyzed directly in the context of functional analysis for networks of Kuramoto oscillators. Stationary solutions of~\eqref{eq:TransportEq} and their stability have been studied recently in the context of all-to-all coupled networks of Kuramoto oscillators~\cite{Mirollo2007,Carrillo2014,Dietert2016,Dietert2016a,Carrillo2019}. Taking the mean-field limit for $\maxdim\to\infty$ depends on the homogeneity of the network. For certain classes of structured networks---networks on convergent families of random where a limiting object (a graphon) can be defined as the number of nodes $\maxdim\to\infty$---it is possible to define and analyze the dynamics of the resulting \emph{continuum limit}~\cite{Medvedev2013a,Chiba2016}.

Ott and Antonsen~\cite{Ott2008} showed that there exists a manifold of invariant probability densities for the transport equation~\eqref{eq:TransportEq}. Specifically, if $f_\sigma(\vth,\omega;0)$ is on the manifold, so will the density $f_\sigma(\vth,\omega;t)$ for any time $t\geq 0$. Let
\begin{align}\label{eq:continuousOP}
 \OP_\sigma &:= \int_{-\infty}^\infty \int_{-\pi}^\pi f_\sigma(\vth,\omega;t)e^{i\vth}\udi\vth \udi\omega
\end{align}
denote the Kuramoto order parameter~\eqref{eq:KuramotoOPFin} in the mean-field limit. We will see below that the evolution on the invariant manifold is now described by a simple ordinary differential equation for~$\OP_\sigma$ for each population~$\sigma$.

In the following we outline the key steps to derive a set of reduced equations and  refer to~\cite{Ott2008,Ott2009,Ott2011} for further details. Let $\bar{w}$ denote the complex conjugate of $w\in\C$. Suppose that~$f_\sigma(\vth,\omega;t)$ can be expanded in a Fourier series in the phase angle~$\vth$ of the form
\begin{align}\label{eq:FourierAnsatz}
 f_\sigma(\vth,\omega;t) &= \frac{h_\sigma(\omega)}{2\pi}\left (
 1 + f^+_\sigma + \bar f^+_\sigma
 \right )
& \text{where} && f^+_\sigma&= \sum_{n=1}^\infty f^{(n)}_{\sigma}(\omega,t) e^{i n\vth}.
\end{align}
Here it is assumed that~$f^+_{\sigma}$ has an analytic continuation into the lower complex half plane $\sset{\Im{\omega}<0}$ (and $f^-_{\sigma}:=\bar f^+_{\sigma}$ into $\sset{\Im{\omega}>0}$); even with this assumption we can solve a large class of problems, but it poses a restriction to a number of practical cases discussed in Section~\ref{sec:Limitations} below.
Ott and Antonsen now imposed the ansatz that Fourier coefficients are \emph{powers of a single function $\alphacoeff_\sigma(\omega,t)$}, 
\begin{align}\label{eq:PowerSeries}
 f^{(n)}_{\sigma}(\omega,t) &= \left(\alphacoeff_\sigma(\omega,t)\right)^n.
\end{align}
If $\abs{\alphacoeff_\sigma(\omega,t)}<1$ this ansatz is equivalent to the Poisson kernel structure for the unit disk, 
$f^+_{\sigma} = (\alphacoeff_\sigma e^{i\vth})/(1-\alphacoeff_\sigma e^{i \vth})$. 
Substitution of~\eqref{eq:FourierAnsatz} into \eqref{eq:TransportEq} yields
\begin{align}\label{eq:alphaEQ}
 \frac{\partial \alphacoeff_\sigma}{\partial t} +i\omega \alphacoeff_\sigma + \frac{1}{2} (H_\sigma\alphacoeff_\sigma^2-\bar{H}_\sigma)&= 0,\
\end{align}
Thus, the ansatz~\eqref{eq:PowerSeries} reduces the integral partial differential equation \eqref{eq:TransportEq} to a single ordinary differential equation 
in $\alphacoeff_\sigma$ for each population~$\sigma$. (More precisely, there is an infinite set of such equations, one for each~$\omega$ with identical structure.)
Finally, with~\eqref{eq:PowerSeries} we obtain
\begin{align}
\label{eq:Zsig}
 \OP_\sigma &= \int_{-\infty}^\infty \bar\alphacoeff_\sigma(\omega,t) h_\sigma(\omega) \udi\omega,\	
\end{align}
which relates~$\alphacoeff_\sigma$ and the order parameter~$\OP_\sigma$ in~\eqref{eq:continuousOP}.

Assuming analyticity, this integral may be evaluated using the residue theorem of complex analysis\footnote{Here we are making an implicit assumption on the regularity of~$H_\sigma$ on time (and potentially an additional parameter): $H_\sigma$ has to be sufficiently smooth that~\eqref{eq:alphaEQ} yields an~$a_\sigma(\omega,t)$ such that the residue theorem can be used to evaluate~\eqref{eq:Zsig} at any time~$t$.}.
These equations take a particularly simple form if the distribution of intrinsic frequencies $h_\sigma(\omega)$ is Lorentzian with mean $\omb_\sigma$ and width $\Delta_\sigma$, i.e.,
\begin{align}
 h_\sigma(\omega)&= \frac{1}{\pi}\frac{\Delta_\sigma}{(\omega-\omb_\sigma)^2+\Delta_\sigma^2},
\end{align}
since $h_\sigma(\omega)$ has two simple poles at $\omb_\sigma\pm i\Delta_\sigma$ and thus~\eqref{eq:Zsig} gives
$\OP_\sigma=\bar\alphacoeff_\sigma(\omb_\sigma-i\Delta_\sigma,t)$
under the assumption $|a_\sigma(\omega,t)|\to 0$ as $\Im{\omega}\to -\infty$.
As a result, we obtain the two-dimensional differential equation---the \textbf{Ott--Antonsen equations} for a Lorentzian frequency distribution---for the order parameter in population~$\sigma$,
\begin{align}\label{eq:OttAnt}\tag{OA}
\dot \OP_\sigma &=  (-\Delta_\sigma+i\omb_\sigma)\OP_\sigma+\frac{1}{2}H_\sigma - \frac{1}{2}\bar H_\sigma{\OP}_\sigma^2.
\end{align}
We note that this reduction method also works for other frequency distributions~$h_\sigma$, as outlined in~\cite{Ott2011}. However, the resulting mean-field equation will not always be a single equation but could be a set of coupled equations. For example, for multi-modal frequency distributions~$h_\sigma$ the Ott--Antonsen equations will have an equation for each mode; see~\cite{Martens2009,Pazo2009,Pietras2016a} and the discussion below. 

The derivation above only states that there \emph{exists} an invariant manifold of densities~$f_\sigma$ for the transport equation~\eqref{eq:TransportEq}. What happens to densities~$f_\sigma$ that are not on the manifold as time evolves? Under some assumptions on the distribution in intrinsic frequencies~$h_\sigma$, Ott and Antonsen also showed in~\cite{Ott2009} that there are densities~$f_\sigma$ that are attracted to the invariant manifold. In other words, the dynamics of the Ott--Antonsen equations capture the long-term dynamics of a wider range of initial phase distributions~$f_\sigma(\vth,\omega;0)$, whether they satisfy~\eqref{eq:PowerSeries} initially or not. We discuss this in more detail below.


\subsubsection{Ott--Antonsen equations for commonly used oscillator models}
\label{sec:OACommonlyUsed}

We now summarize the Ott--Antonsen equations~\eqref{eq:OttAnt} for the commonly used oscillator models described in the Section~\ref{sec:OscillatorModels}. Here we focus on Lorentzian distributions of the intrinsic frequencies or excitabilities; for Ott--Antonsen equations for other parameter distributions such as normal or bimodal distributions see~\cite{Ott2008,Martens2009}.

\paragraph{The Kuramoto model.}
Consider the mean-field limit of the Kuramoto model~\eqref{eq:Kuramoto} with a Lorentzian distribution of intrinsic frequencies. Recall that the driving field for the Kuramoto model was $H(t)=K\OP(t)$. Substituting this into~\eqref{eq:OttAnt} we obtain \emph{Ott--Antonsen equations for the Kuramoto model}
\begin{align}\label{eq:KuramotoMF}
\dot \OP &=  (-\Delta+i\omb)\OP+\frac{K}{2}\OP\left(1 - \abs{\OP}^2 \right),
\end{align}
a two-dimensional system of equations since~$Z$ is complex-valued.

\paragraph{Kuramoto--Sakaguchi equations.}
For the Kuramoto--Sakaguchi equations~\eqref{eq:PopKuramotoSakaguchi}
the driving field is a weighted sum of the individual population order parameters~\eqref{eq:DriveKS}. Assuming a Lorentzian distribution of intrinsic frequencies with mean~$\omb_\sigma$ and width~$\Delta_\sigma$ for each population  $\sigma\in\sset{1,\ldots,\maxpop}$, we obtain from~\eqref{eq:OttAnt} the \emph{Ott--Antonsen equations for coupled populations of Kuramoto--Sakaguchi oscillators},
\begin{equation}\label{eq:OttAntMultiPop}
  \dot \OP_\sigma = (-\Delta_\sigma+i\omb_\sigma)\OP_\sigma + \half\left(\sum_{\tau=1}^\maxpop c_{\sigma\tau}\OP_\tau-\OP_\sigma^2\sum_{\tau=1}^\maxpop \bar{c}_{\sigma\tau}\bar{\OP}_\tau\right).
\end{equation}
In other words, the Ott--Antonsen equations are a $2\maxpop$-dimensional system that describe the interactions of the order parameters~$\OP_\sigma$.

\paragraph{Networks of Theta neurons.}
Consider a single population of Theta neurons with drive~$I(t)$ given by~\eqref{eq:dthdtD} with parameter-dependent intrinsic frequencies and driving field~\eqref{eq:DriveTheta}; we omit the population index~$\sigma$. Assume that the variations in excitability~$\eta_k$ are chosen from a Lorentzian distribution mean~$\etb$ and width~$\Delta$. We obtain the \emph{Ott--Antonsen equations for the mean-field limit of a population of Theta neurons~\eqref{eq:DriveTheta}}
\be
   \dot \OP=\frac{1}{2}\left((i\etb-\Delta)(1+\OP)^2-i(1-\OP)^2\right)+\frac{1}{2}i(1+\OP)^2\kappa I.\label{eq:dzdtA}
\ee
Note that in contrast to~\eqref{eq:OttAntMultiPop}, this is not a closed set of equations yet as the exact form of the input current is still unspecified. We will close these equations in Section~\ref{sec:ThetaNeurons} below by writing~$I$ in terms of~$\OP$ for different types of neural interactions.

The order parameter for the Theta neuron directly relates to quantities with a physical interpretation such as the average firing rate of the network. Integrating the phase distribution~\eqref{eq:FourierAnsatz} over the excitability parameter~$\eta$ under assumption~\eqref{eq:PowerSeries}
we obtain the distribution of all phases,
\be
   p(\theta,t)=\frac{1}{2\pi}\left(\frac{1-\abs{\OP}^2}{1-\OP e^{-i\theta}-\bar{\OP}e^{i\theta}+\abs{\OP}^2}\right)
=\frac{1}{2\pi}\Re{\frac{1+\bar{\OP}e^{i\theta}}{1-\bar{\OP}e^{i\theta}}}, \label{eq:pdist}
\ee
where~$\OP$ may be a function of time. This distribution can be used to determine the probability that a Theta neuron has phase~$\theta$. Since a Theta neuron fires when its phase crosses $\theta=\pi$, the average firing rate~$r(t)$ of the network at time~$t$ is the flux through $\theta=\pi$, i.e.,
\be\label{eq:firingrate1}
   r(t)=(p(\theta,t)\dot{\theta})|_{\theta=\pi}=\frac{1}{\pi}\Re{\frac{1-\bar{\OP}(t)}{1+\bar{\OP}(t)}}.
\ee
Here we used that $\dot{\theta}|_{\theta=\pi}=2$ by~\eqref{eq:dthdtD}, independent of~$\theta$. The same result is obtained from the firing rate equations of the QIF neuron as we explain in the next paragraph.

\subsubsection{Ott--Antonsen reduction for equivalent networks}\label{sec:FRE}

The mean-field reductions are also valid for systems that are \emph{equivalent} to a network of sinusoidally coupled phase oscillators~\eqref{eq:Microscopic}. As an example, we discussed the relationship between QIF and Theta neurons above via the transformation $V=\tan{(\theta/2)}$, which carries over to the mean-field limit of infinitely many neurons where the Ott--Antonsen equations apply. More specifically, this transformation converts the distribution of phases~\eqref{eq:pdist} into a distribution
\be
   \tilde{p}(V,t)=\frac{X(t)}{\pi((V-Y(t))^2+X^2(t))} \label{eq:Vdist}
\ee
of voltages where $Z = (1-\bar{W})/(1+\bar{W})$ and $W=X+iY$ and $X,Y\in\mathbb{R}$. Equation~\eqref{eq:Vdist} is called the \emph{Lorentzian ansatz} in~\cite{Montbrio2015}. Importantly, the quantity~$W$ is obtained from a conformal transformation of the order parameter~$\OP$. This allows one to convert the Ott--Antonsen equations for the Theta neurons~\eqref{eq:dzdtA} to an equation for the mean field~$W=(1-\bar{\OP})/(1+\bar{\OP})$, given by
\be
   \dot W = i\etb+\Delta-iW^2+iI, \label{eq:dwdt}
\ee
which describes the QIF neurons. The advantage of this formulation is that both the real and imaginary parts of~$W$ have physical interpretations: $Y(t)$ is the average voltage across the network and $X(t)$ relates to the firing rate~$r$ of the population, i.e.,~the flux at $V=\infty$, since $\lim_{V\rightarrow\infty}\tilde{p}(V,t)\dot V(t) = X(t)/\pi=r$~\cite{Montbrio2015}.

\subsection{Watanabe--Strogatz reduction for identical oscillators}
\label{sec:RedWS}

Mean-field reductions are possible for both finite and infinite networks for populations of identical oscillators. These reductions are due to the high level of degeneracy in the system, i.e.,~there are many quantities that are conserved as time evolves. This degeneracy was first observed in the early 1990s for coupled Josephson junction arrays~\cite{Tsang1991}, which relate directly to Kuramoto's model of coupled phase oscillators~\cite{Wiesenfeld1998}. Watanabe and Strogatz~\cite{Watanabe1993,Watanabe1994} were able to calculate the preserved quantities explicitly using a clever transformation of the phase variables, thereby reducing the Kuramoto model from $\maxdim$ (oscillator phases) to three time-dependent (mean-field) variables together with $\maxdim-3$ constants of motion. In terms of mathematical theory, the degeneracy originates from restrictions imposed by the algebraic structure of the equations~\cite{Goebel1995,Marvel2009,Stewart2011} which is still an area of active research~\cite{Chen2017,Engelbrecht2020}. 

The Watanabe--Strogatz reduction applies for sinusoidally coupled phase oscillator populations where oscillators within populations are identical, i.e., all oscillators have the same intrinsic frequency, $\omega_{\sigma,k} = \omega_{\sigma}$, and are driven by the same field $H_{\sigma,k} = H_\sigma$. Indeed, Watanabe--Strogatz and Ott--Antonsen reductions have been shown to be intricately linked~\cite{Marvel2009,Pikovsky2011} as we briefly discuss below. Here, we focus on finite networks for simplicity. In the following section we give the equations in generality and give some mathematical detail. Then, the equations are subsequently stated for the commonly used oscillator models discussed above.

\subsubsection{*Constants of motion yield reduced equations}

\begin{figure}
\begin{center}
\includegraphics[width=0.25\textwidth]{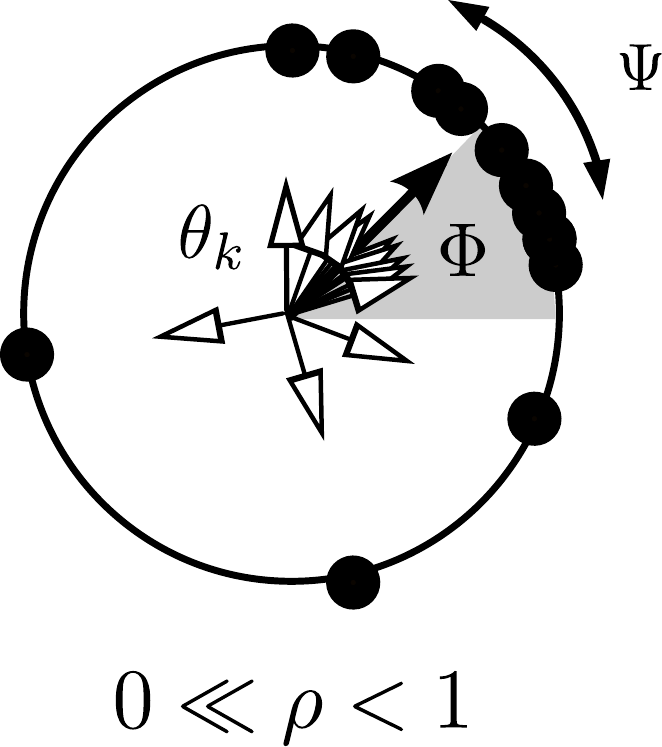}
\smallskip
\end{center}
\caption{\label{fig:BunchVars}Illustration of the bunch variables in the Watanabe--Strogatz formalism. 
Just like the Kuramoto order parameter~$\OP_\sigma$, the bunch amplitude and bunch phase in~$z_\sigma=\rho_\sigma e^{i \Phi_\sigma}$ characterize the level of synchrony. The quantities~$\OP_\sigma$ and~$z_\sigma$ do however only coincide if the population is fully synchronized or  for uniformly distributed constants of motion in the limit $N\rightarrow \infty$ (see text). The phase distribution variable~$\Psi_\sigma$ is related to the shift and distribution of individual oscillators with respect to~$\Phi_\sigma$.
}
\end{figure}

The dynamics of a finite population~\eqref{eq:Microscopic} with $\maxdim>3$ identical oscillators can be described exactly in terms of three macroscopic (mean-field) variables~\cite{Watanabe1993,Watanabe1994,Pikovsky2008,Pikovsky2011}: the bunch amplitude~$\rho_\sigma$, bunch phase~$\Phi_\sigma$, and phase distribution variable~$\Psi_\sigma$.
Similar to the modulus and phase of the Kuramoto order parameter~$\OP_\sigma=R_\sigma e^{i\phi_\sigma}$, the bunch amplitude~$\rho_\sigma$ and bunch phase~$\Phi_\sigma$ characterize synchrony (or equivalently, the maximum of the phase distribution); while~$(R_\sigma, \phi_\sigma)$ and $(\rho_\sigma,\Phi_\sigma)$ do not coincide in general, they do if the population is fully synchronized. The phase distribution variable~$\Psi_\sigma$ determines the shift of individual oscillators with respect to~$\Phi_\sigma$ as illustrated in Fig.~\ref{fig:BunchVars}.

For a population of sinusoidally coupled phase oscillators~\eqref{eq:Microscopic} with driving field $H_\sigma=H_\sigma(t)$ the macroscopic variables evolve according to the \textbf{Watanabe--Strogatz equations}
{\allowdisplaybreaks
\begin{subequations}
\begin{align}
  \label{eq:WSrh}\tag{WSa}\dot\rho_\sigma &= \frac{1-\rho_\sigma^2}{2}\Re{H_\sigma e^{-i\Phi_\sigma}},\\
  \label{eq:WSPh}\tag{WSb}\dot\Phi_\sigma &= \omega_\sigma+\frac{1+\rho_\sigma^2}{2\rho_\sigma}\Im{H_\sigma e^{-i\Phi_\sigma}},\\
  \label{eq:WSPs}\tag{WSc}\dot\Psi_\sigma &= \frac{1-\rho_\sigma^2}{2\rho_\sigma}\Im{H_\sigma e^{-i\Phi_\sigma}}.
\end{align}
\end{subequations}}%
Mathematically speaking, the reduction to three variables means that the phase space~$\Torn$ of~\eqref{eq:Microscopic} is foliated by 3-dimensional leafs, each of which is determined by constants of motion,~$\psiCM{\sigma}{k}$, $k=1, \dotsc, \maxdim$ ($\maxdim-3$ are independent). In other words, the choice of constants of motion determines a specific $3$-dimensional invariant subspace on which the macroscopic variables evolve. The Watanabe--Strogatz equations arise from the properties of Riccati equations and the bunch variables are parameters of a family of M\"obius transformations which determine the system's dynamics; see~\cite{Marvel2009,Stewart2011,Chen2017,Engelbrecht2020} for more details on the mathematics behind these equations.

From a practical point of view, two things are needed to use the Watanabe--Strogatz equations~\WS to understand oscillator networks of the form~\eqref{eq:Microscopic}. First, since the driving field~$H$ is often a function of the population order parameters~$\OP_\tau$, $\tau=1,\ldots,\maxpop$, we need to translate~$\OP_\sigma$ into the bunch variables to get a closed set of equations. Write $z_\sigma:=\rho_\sigma e^{i\Phi_\sigma}$. As shown for example in~\cite{Pikovsky2011}, we have
\begin{align}\label{eq:Bunch2Z}
  \OP_\sigma &= \BP_\sigma\gamma_\sigma &\text{where}&& \gamma_\sigma(\rho_\sigma,\Psi_\sigma) &= \frac{1}{\maxdim\rho_\sigma}\sum_{j=1}^\maxdim\frac{\rho_\sigma e^{i\Psi_\sigma} + e^{i\psiCM{\sigma}{j}}}{e^{i\Psi_\sigma} + \rho_\sigma e^{i\psiCM{\sigma}{j}}}.
\end{align}
Second, one needs to determine the constants of motion from the initial phases $\theta_{\sigma,k}(0)$: A possible choice is to set $\psiCM{\sigma}{k}:=\theta_{\sigma,k}(0)$ and $\rho_\sigma(0)=\Phi_\sigma(0)=\Psi_\sigma(0)=0$; see~\cite{Watanabe1994} for a detailed discussion and different way to choose initial conditions that avoids the singularity at $\rho_\sigma=0$. 
Taken together, the dynamics of individual oscillators~\eqref{eq:Microscopic} are now determined by~\WS via~\eqref{eq:Bunch2Z} and vice versa.

The relationship~\eqref{eq:Bunch2Z} between the bunch variables and the order parameter also indicates how the Watanabe--Strogatz equations and the Ott--Antonsen equations are linked. Pikovsky and Rosenblum~\cite{Pikovsky2008} showed that for constants of motion that are uniformly distributed on the circle, $\psiCM{\sigma}{k} = 2\pi k/\maxdim$, we have $\gamma_\sigma \to 1$  as $\maxdim\to\infty$.
Consequently, $Z_\sigma=z_\sigma$ for such a choice of constants of motion in the limit of infinitely many oscillators.
For the Kuramoto model with $H_\sigma = Z_\sigma$, equations~\eqref{eq:WSrh} and~\eqref{eq:WSPh} depend on~$\Psi$ only through~$\gamma$. 
Thus, for constant $\gamma=1$ the equations~\eqref{eq:WSrh} and~\eqref{eq:WSPh} decouple from~\eqref{eq:WSPs}. These two equations are equivalent to the Ott--Antonsen equations~\eqref{eq:OttAnt} in the mean-field limit for identical oscillators. To summarize, the dynamics of the mean-field limit for identical oscillators is given by the Watanabe--Strogatz equations together with a \emph{distribution} of constants of motion. For the particular choice of a uniform distribution of constants of motion, the equations decouple and the effective dynamics are given by the Ott--Antonsen equations.

\subsubsection{Watanabe--Strogatz equations for commonly used oscillator models}
\label{sec:WSCommonlyUsed}

We now summarize the Watanabe--Strogatz equations~\WS for the commonly used oscillator models described in Section~\ref{sec:OscillatorModels}.

\paragraph{Kuramoto--Sakaguchi equations.}
For the multi-population Kuramoto--Sakaguchi model~\eqref{eq:PopKuramotoSakaguchi}, the driving field~$H$ is a linear combination of the order parameters, $H_\sigma = \sum_{\tau=1}^\maxpop c_{\sigma\tau}\OP_\tau$. Assuming that the oscillators within each population are identical, $\omega_{\sigma,k}=\omega_\sigma$, the dynamics are governed by the \emph{Watanabe--Strogatz equations for coupled Kuramoto--Sakaguchi populations}
{\allowdisplaybreaks
\begin{subequations}\label{eq:WS_KS}
\begin{align}
  \label{eq:WSrhKS}\dot\rho_\sigma &= \frac{1-\rho_\sigma^2}{2}\Re{\sum_{\tau=1}^\maxpop c_{\sigma\tau}\gamma_{\tau}\rho_{\tau} e^{i(\Phi_\tau-\Phi_\sigma)}},\\
  \label{eq:WSPhKS}\dot\Phi_\sigma &= \omega_\sigma+\frac{1+\rho_\sigma^2}{2\rho_\sigma}\Im{\sum_{\tau=1}^\maxpop c_{\sigma\tau}\gamma_{\tau}\rho_{\tau} e^{i(\Phi_\tau-\Phi_\sigma)}},\\
  \label{eq:WSPsKS}\dot\Psi_\sigma &= \frac{1-\rho_\sigma^2}{2\rho_\sigma}\Im{\sum_{\tau=1}^\maxpop c_{\sigma\tau}\gamma_{\tau}\rho_{\tau} e^{i(\Phi_\tau-\Phi_\sigma)}}.
\end{align}
\end{subequations}}%

\paragraph{Networks of Theta neurons.}
For a finite population of identical Theta neurons~\eqref{eq:DriveTheta} with identical excitability~$\eta$ and input current~$I(t)$ the \emph{Watanabe--Strogatz equations for identical Theta neurons}~\cite{laing2018} evaluate to
{\allowdisplaybreaks
\begin{subequations}\label{eq:WSth}
\begin{align}
  \label{eq:WSrhth}\dot\rho &= \frac{1-\rho^2}{2}\Re{i(\eta+\kappa I-1)e^{-i\Phi}},\\
  \label{eq:WSPhth}\dot\Phi &= 1+\eta+\kappa +\frac{1+\rho^2}{2\rho}\Im{i(\eta+\kappa I-1)e^{-i\Phi}},\\
  \label{eq:WSPsth}\dot\Psi &= \frac{1-\rho^2}{2\rho}\Im{i(\eta+\kappa I-1)e^{-i\Phi}}.
\end{align}
\end{subequations}}%
Note that, as for the Ott--Antonsen reduction above, one still needs to close this system by writing~$I$ in terms of the bunch variables in~\WS and the constants of motion. This is not straightforward and requires a considerable amount of computations~\cite{laing2018}.

\subsubsection{Reductions for equivalent networks}
For a finite network of identical QIF neurons governed by~\eqref{eq:QIF} with $\eta_j=\eta$ for all $j$, the transformation $V=\tan{(\theta/2)}$ converts this network into a network of identical Theta neurons~\eqref{eq:dthdtD}. Consequently, such a network will also be described by equations of the form~\eqref{eq:WSth}. As mentioned above, in the limit $\maxdim\rightarrow\infty$ and equally spaced
constants of motion, the equation~\eqref{eq:WSPsth} will decouple from~\eqref{eq:WSrhth} and~\eqref{eq:WSPhth}.
In this case, writing $z=\rho e^{i\Phi}$ we find that~$z$ satisfies~\eqref{eq:dzdtA} or equivalently~\eqref{eq:dwdt} (with $\etb=\eta$ and $\Delta=0$).

\subsection{Limitations and challenges}
\label{sec:Limitations}

Before we apply the mean-field reductions to particular oscillator networks in the next section, some (mathematical) comments on the limitations of these approaches are in order.

The main assumption behind the reduction methods is that network interactions are mediated by a coupling function with a single harmonic (of arbitrary order). There are explicit examples~\cite{Bick2011,Lai2013a,Bick2016b} that show that the reductions, as described above, become invalid. For example chaotic dynamics may occur where the reduction would have an effective two-dimensional phase space; we discuss this example below. 
This does not mean that the reductions break down completely, and there may still be some degeneracy in the system if the interaction is of a specific form; see~\cite{Ashwin2016a} for a more detailed discussion. It remains a challenge to identify what part of the mean-field reduction (if any) remains valid for more general interaction functions and phase response curves.

The Ott--Antonsen reduction for the mean-field limit allows for the oscillators to be nonidentical. By contrast, the Watanabe--Strogatz reduction of finite networks requires oscillators to be identical. Neither of these approaches applies to finite networks of nonidentical oscillators, and understanding such networks remains a challenge.
Direct numerical simulations to elucidate the dynamics of networks of~$\maxdim$ almost identical oscillators are challenging as one needs to integrate an almost integrable dynamical system\footnote{This problem can however be solved by respecting the symmetries underlying the system, i.e., either by integrating the WS equations or the dynamic equations governing the Möbius transformations, which in turn can be used to compute trajectories for individual oscillators.}.
There has also been some recent progress analyzing situations in which the Ott--Antonsen or Watanabe--Strogatz equations do not apply.
First, a perturbation theory for the exact mean-field equations has been developed to elucidate the dynamics for systems that are close to sinusoidally coupled, for example if there are very weak higher-harmonics in the interaction function~\cite{Vlasov2016}.
Second, while not an exact representation of the dynamics, the collective coordinates approach by Gottwald and coworkers~\cite{Gottwald2015,Gottwald2017,Smith2019} has been instructive to gain insights into the dynamics of finite networks of nonidentical oscillators.

Finally, Ott and Antonsen showed that the manifold of oscillator densities~$f_\sigma$ on which the reduction holds is attracting~\cite{Ott2009}. Their method of proof has been shown to apply to a wider class of systems~\cite{Pietras2016a}. As pointed out by Mirollo~\cite{Mirollo2012} and later elaborated further~\cite{Engelbrecht2020}, their proof is based on a strong smoothness assumption on the density~$f_\sigma$ which implies limitations to this approach. More precisely, to be able to evaluate contour integrals using the residue theorem, it is typically assumed that the integrand in~\eqref{eq:Zsig}, containing the intrinsic frequency distribution~$h_\sigma$ and the density~$f_\sigma$, is holomorphic. In particular, this assumption is only valid for distributions~$h_\sigma$ that allow for arbitrarily large (or small) intrinsic frequencies with nonzero probability:
The identity theorem for holomorphic functions implies that $h_\sigma(\omega)>0$ for all~$\omega\in\R$. Any distribution for which the intrinsic frequencies are bound to a finite interval---the intrinsic frequencies of any finite collection of oscillators will lie in a finite interval---are excluded\footnote{Compactly supported distributions of intrinsic frequencies have been approximated by rational distributions~\cite{Skardal2018,Pietras2018}, but it is not clear whether the limit is independent of the approximation.}. Hence, while the manifold described by Ott and Antonsen attracts some class of oscillator densities, it is not clear how large this class actually is (it does not include $\delta$-distributions where all oscillators have the same phase). Put differently, it is important to explicitly characterize the space of densities in which the Ott--Antonsen manifold is attracting.

\section{Dynamics of coupled oscillator networks}
\label{sec:UsingMFR}

We now discuss global synchrony and synchrony patterns in phase oscillator networks, and  highlight how the reductions presented in the previous section simplify their analysis. While we indicate along the way how most of these systems are relevant from the point of view of biology and neuroscience, we here take a predominantly dynamical systems perspective and highlight the applicability of, for example, bifurcation theory~\cite{Strogatz1994book,Kuznetsov2004}. We focus on a small number of coupled populations of oscillators, which can be seen as building blocks for larger models consisting of many coupled populations (e.g., regions of interest in a whole-brain model as discussed in Section~\ref{sec:NeuroModel} below).

\subsection{Networks of Kuramoto-type oscillators}

We first consider networks of Kuramoto--Sakaguchi and related Kuramoto-type oscillators. Despite their simplicity, they have found widespread application, for example in neuroscience, as outlined in Section~\ref{sec:KuramSakaguchi}, to understand synchronization phenomena. The network interactions of such oscillators depend on phase differences. Bifurcations may occur as one introduces an explicit phase dependency to the coupling~\cite{Brown2003} such as in the networks of Theta neurons which we discuss in the following section.

\subsubsection{One oscillator population}

\noindent\textbf{Example 1:\ }\label{example1}%
We first revisit Kuramoto's original problem (see \hyperref[problem1]{\emph{Problem~1}} in Section~\ref{sec:KuramotoModel} above) from the perspective of mean-field reductions: 
Given a globally coupled network of Kuramoto oscillators~\eqref{eq:Kuramoto} with a Lorentzian  distribution of intrinsic frequencies, what is the critical coupling strength~$K_c$ where oscillators start to synchronize? 

This problem is surprisingly easy to solve in the mean-field limit $\maxdim\to\infty$ using the Ott--Antonsen reduction. Assume that the distribution of intrinsic frequencies is a Lorentzian with mean~$\omb$ and width~$\Delta$. Recall that the order parameter~$\OP$ evolves according to the Ott--Antonsen equation~\eqref{eq:KuramotoMF}: Separating~\eqref{eq:KuramotoMF} for $\OP = Re^{i\phi}$ into real and imaginary parts yields%
\begin{subequations}%
\begin{align}
\dot R &= \left(-\Delta+\frac{K}{2}-\frac{K}{2}R^2 \right)R,\label{eq:KuramotoMFOP}\\
\dot \phi &= \omb.\label{eq:KuramotoMFOPb}
\end{align}
\end{subequations}
Moreover, the manifold on which~\eqref{eq:KuramotoMF} describes the mean-field limit of~\eqref{eq:Kuramoto} attracts initial phase distributions. 
Since the equation for the mean phase~$\phi$ is completely uncoupled, it suffices to analyze~\eqref{eq:KuramotoMFOP}. Thus, Kuramoto's problem in the infinite-dimensional mean-field-limit reduces to solving the one-dimensional real ordinary differential equation~\eqref{eq:KuramotoMFOP}: By elementary analysis, we find that the equilibrium $R=0$ is stable for $K<K_c=2\Delta$ and loses stability in a pitchfork bifurcation where the solution $R  =\sqrt{1-2\Delta/K}>0$ becomes stable. The same analysis applies to the Kuramoto--Sakaguchi network~\eqref{eq:PopKuramotoSakaguchi} with $\maxpop=1$ for phase-lag~$\alpha\in\big(-\frac{\pi}{2}, \frac{\pi}{2}\big)$ with~$K$ replaced by $K\cos(\alpha)$ in~\eqref{eq:KuramotoMF}; note that for phase lag $\sin{\alpha}\neq 0$ we have~$\dot\phi=\omb+K\sin{(\alpha)}R(1-R^2)$ so that the frequency now depends nontrivially on~$R$.

Global synchronization of finite networks of identical Kuramoto--Sakaguchi oscillators is readily analyzed using the Watanabe--Strogatz reduction. As above, a phase variable decouples and we obtain a two-dimensional system which describes the dynamics of~\eqref{eq:PopKuramotoSakaguchi} for $\maxpop=1$. Its analysis~\cite{Watanabe1993} shows that the system will synchronize perfectly, $R\to1$ as $t\to\infty$, for $\alpha\in\big(-\frac{\pi}{2}, \frac{\pi}{2}\big)$ (attractive coupling) and converge to an incoherent equilibrium, $R\to0$ as $t\to\infty$, for $\alpha\in\big(\frac{\pi}{2}, \frac{3\pi}{2}\big)$ (repulsive coupling). In the marginal case of $\cos(\alpha) = 0$ the system is Hamiltonian~\cite{Watanabe1994}.

\paragraph{Multimodal distributions in the Kuramoto model.}
While Kuramoto's original model considered a single oscillator population with unimodally distributed frequencies---such as the Lorentzian distribution---Kuramoto also speculated on what dynamic behaviors 
a network consisting of a single population
would exhibit if the distribution of natural frequencies was instead bimodal~\cite{Kuramoto1984}: Depending on the coupling strength, the width and spacing of the peaks of the frequency distribution, oscillators may either aggregate and form a single crowd of oscillators, thus forming one ``giant oscillator,'' or disintegrate into two mutually unlocked crowds, corresponding to two giant oscillators.

Crawford analyzed this case rigorously for the weakly nonlinear behavior near the incoherent state using center manifold theory~\cite{Crawford1994} and thus explained local bifurcations in the neighborhood of the incoherent state. Using the Ott--Antonsen reduction, Martens {\it et al.}~\cite{Martens2009} obtained exact results on all possible bifurcations and the bistability between incoherent, partially synchronized, and traveling wave solutions. Similarly, rather than superimposing two unimodal frequency distributions, Paz{\'o} and Montbri\'{o}~\cite{Pazo2009} considered a modified model where the distribution of intrinsic frequencies is the difference of two Lorentzians; this allows for the central dip to become zero%
\footnote{The resulting frequency distribution is curiously similar to Norbert Wiener's notion of the frequency distribution of brain waves around the alpha wave band, see, e.g.,~\cite{Strogatz1994}.}.

Interestingly, to describe a single population with an $m$-modal frequency distribution using the Ott--Antonsen reduction, one obtains a set of~$m$ coupled ordinary differential equations. This set describes the oscillator dynamics of~$m$~order parameters \eqref{eq:continuousOP} associated with each peak of the $m$-modes, resulting in collective behavior where oscillators either aggregate to a single or potentially up to~$m$ groups of oscillators.
The question arises as to whether the resulting set of equations can be related to $\maxpop$-population models as described by~\eqref{eq:PopKuramotoSakaguchi}. This question was picked up by Pietras and Daffertshofer~\cite{Pietras2016} who showed that the dynamical equations describing $\maxpop=1$ population with a bimodal distribution can be mapped to $\maxpop=2$ populations~\eqref{eq:PopKuramotoSakaguchi} with nonidentical coupling strengths~$K_{\sigma\tau}$ with equivalent bifurcations. However, this equivalence breaks down for $\maxpop=3$ populations and trimodal distributions.

\paragraph{Higher-order and nonadditive interactions.}
Note that networks of Kuramoto--Sakaguchi oscillators~\eqref{eq:PopKuramotoSakaguchi} make two important assumptions on the network interactions. First, the interactions are sinusoidal, as discussed above, since the coupling function has a single harmonic. Second, the network interactions are \emph{additive}~\cite{Bick2015,Aguiar2018}, that is, the interaction of two distinct oscillators on a third is given by the sum of the individual interactions. By contrast, coupling between oscillatory units generically contains nonlinear (nonadditive) interactions; concrete examples include oscillator networks~\cite{Tanaka2011a}, interactions in ecological networks~\cite{Levine2017}, and nonlinear dendritic interactions between neurons~\cite{Ariav2003,Polsky2004,Memmesheimer2010}.
For weakly coupled oscillator networks, \emph{higher-order} interaction terms include higher harmonics in the coupling function as well as coupling terms which depend nonlinearly on three or more oscillator phases~\cite{Rosenblum2007}. Such terms naturally arise in phase reductions: If the interaction between the nonlinear oscillators is generic, Ashwin and Rodrigues~\cite{Ashwin2015a} calculated these corresponding higher-order interaction terms explicitly for a globally coupled network of symmetric oscillators close to a Hopf bifurcation. 
Moreover, higher-order interactions in the effective phase dynamics can also arise for additively coupled nonlinear oscillators~\cite{Kralemann2014}, for example in higher-order phase reductions~\cite{Leon2019a}. Nonadditive interactions can be exploited for applications, such as to build neurocomputers~\cite{Hoppensteadt1999}.

The mean-field reductions here can be used to analyze networks with particular type of higher-order interactions. For example, Skardal and Arenas~\cite{Skardal2019a} consider a single globally coupled population of indistinguishable oscillators where the pure triplet interactions of the form $\sin(\theta_l+\theta_j-2\theta_k)$ determines the joint influence of oscillators~$j,l$ onto oscillator~$k$. In the mean-field limit, they find multistability and hysteresis between incoherent and partially synchronized attractors.
In general, however, higher-order interaction terms lead to phase oscillator networks where the mean-field reductions cease to apply~\cite{Bick2016b}.

\paragraph{Generalizations.}
Much progress has been made to understand synchronization and more complicated collective dynamics in globally coupled networks of Kuramoto oscillators and their generalizations; see~\cite{Brown2003,Acebron2005,Pikovsky2015a} for surveys. While we discussed Kuramoto's problem as an example, the same methods apply for more general types of driving fields~$H$: They may include homogeneous~\cite{Ott2008} or heterogeneous delays~\cite{Lee2009,Petkoski2016} (the latter one being of specific interest for coupled populations of neurons), they may be heterogeneous in terms of the contribution of individual oscillators~\cite{Lohe2017}, or they may include generalized mean fields~\cite{Chen2017}. However, note that much richer dynamics are possible when the assumptions of sinusoidal coupling breaks down. Because of the Poincar\'e--Bendixson theorem~\cite{Schwartz1963,Strogatz1994book}, chaos is not possible for the mean-field reductions for $\maxpop=1$ populations of Kuramoto--Sakaguchi oscillators since their effective dynamics is one- or two-dimensional, respectively. By contrast, even for fully symmetric networks, higher harmonics in the phase response curve/coupling function may lead to chaotic dynamics~\cite{Bick2011,Bick2016b}.

\begin{figure}
\centering
\includegraphics[width=\scl\textwidth]{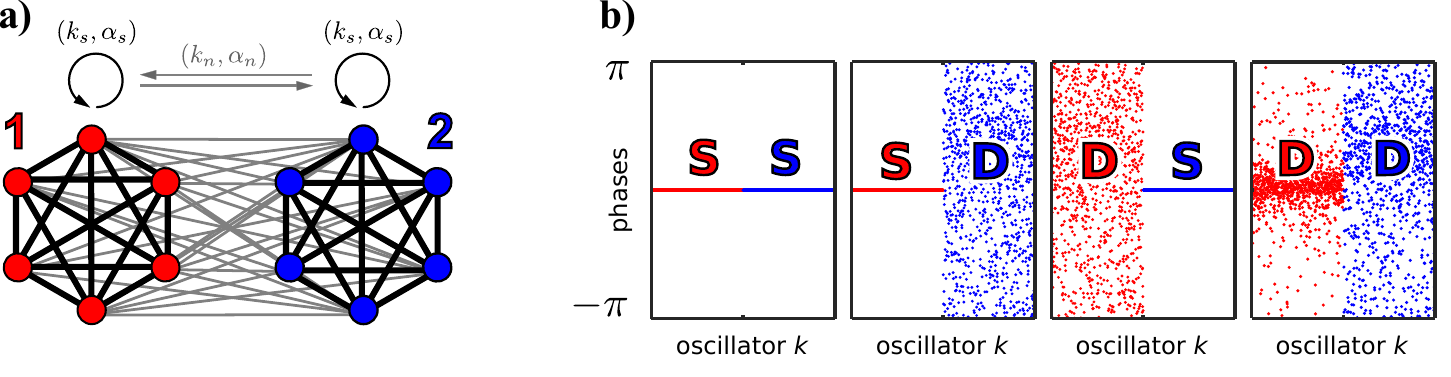}
\medskip
\caption{Synchrony patterns arise in networks of $\maxpop=2$ populations. Panel~(a) shows a cartoon of the network structure; the nodes of population~$\sigma=1$ are colored in red and the nodes of population $\sigma=2$ in blue. The coupling within each populations (black edges) is determined by the coupling strength~$k_s$ and phase lag~$\alpha_s$ and between populations (gray edges) by~$k_n$ and~$\alpha_n$. Panel~(b) shows various stable synchrony patterns in the network as phase snapshots of the solutions to Eqs.~\eqref{eq:PopKuramotoSakaguchi} with $\maxdim=1000$ oscillators per population once the system has relaxed to an attractor; here~$\Sp$ indicates a population that is (fully) phase synchronized ($R_\sigma = 1$) and~$\Dp$ a nonsynchronized population ($R_\sigma < 1$). The parameters are $\alpha_s = 1.58$ in the first three plots---here different initial conditions converge to different attractors---and $\alpha_s=1.64$ in the rightmost. The parameters $A=0.7,\alpha_n=0.44$ were the same in all plots.
\label{fig:PATTERNSM2}}
\end{figure}

\subsubsection{Two oscillator populations}\label{sec:2PopPhsLg}

Two coupled populations of Kuramoto--Sakaguchi oscillators can give rise to a larger variety of synchrony patterns. Before considering general coupling between populations, we first discuss the widely investigated case of identical (and almost identical) populations of Kuramoto--Sakaguchi oscillators~\eqref{eq:PopKuramotoSakaguchi} with Lorentzian distribution of intrinsic frequencies. To be precise, we say that all \emph{populations of~\eqref{eq:PopKuramotoSakaguchi} are identical} if for any two populations~$\sigma$, $\tau$, there is a permutation which sends~$\sigma$ to~$\tau$ and leaves the corresponding equations~\eqref{eq:OttAntMultiPop} for the mean-field limit invariant. Intuitively speaking, this means we can swap any population with any other population without changing the dynamics. Mathematically speaking, the populations are identical if the Ott--Antonsen equations~\eqref{eq:OttAntMultiPop} have a permutational symmetry group that acts transitively~\cite{Golubitsky2002}. Note that for the populations to be identical, the oscillators do not need to be identical. But if the populations are identical, then the frequency distributions~$h_\sigma$ are the same for all populations. Moreover, if the oscillators within each population have the same intrinsic frequency (as required for the Watanabe--Strogatz reduction) then all oscillators in the network have the same intrinsic frequency.

Oscillator networks which are organized into distinct populations support \emph{synchrony patterns} which may be \emph{localized}, that is, some populations show more (or less) synchrony than others. While this may not be surprising if the populations are nonidentical, such dynamics may also occur when the populations are identical. For identical populations of Kuramoto--Sakaguchi oscillators, the localized dynamics arise purely through the network interactions---the populations would behave identically if uncoupled---and hence constitute a form of dynamical symmetry breaking. The phenomenon of ``coexisting coherence and incoherence'' has been dubbed a \emph{chimera state} in the literature~\cite{Abrams2004} and has attracted a tremendous amount of attention in the last two decades; see~\cite{PanaggioAbramsReview2015,Scholl2016,Omelchenko2018} for recent reviews. To date, an entire zoo of chimeras and chimera-like creatures has emerged in a range of different networked dynamical systems---with attempts to classify and distinguish these creatures~\cite{Kemeth2016,Kemeth2018}---beyond the original context of phase oscillators~\cite{Kuramoto2002}. Here we will discuss chimeras only in coupled populations of Kuramoto--Sakaguchi oscillators~\eqref{eq:PopKuramotoSakaguchi} as examples of localized patterns of (phase and frequency) synchrony.

\paragraph{Synchrony patterns for two identical populations.}
The Ott--Antonsen reduction has been instrumental to understand the dynamics of networks consisting of $\maxpop=2$ populations of Kuramoto--Sakaguchi oscillators. Assuming that all intrinsic frequencies are distributed according to a Lorentzian, we obtain two coupled Ott--Antonsen equations~\eqref{eq:OttAntMultiPop} for the limit of infinitely large populations. In this section we focus on networks of identical populations, that is, the distributions of intrinsic frequencies are the same and coupling is symmetric; cf.~Fig.~\ref{fig:PATTERNSM2}(a). This allows one to simplify the parametrization of the system by introducing \emph{self-coupling} $c_s = k_s e^{-i\alpha_s} := c_{11} = c_{22}$ and \emph{neighbor-coupling} $c_n = k_n e^{-i\alpha_n} := c_{12} = c_{21}$ parameters and the coupling strength disparity $A=(k_s-k_n)/(k_s+k_n)$.  Writing $Z_\sigma=R_\sigma e^{i\phi_\sigma}$ as above, the state of~\eqref{eq:OttAntMultiPop} is fully determined by the amount of synchrony in each population~$R_1$, $R_2$ and the difference of the mean phase $\psi:=\phi_1-\phi_2$ of the two populations; cf.~\cite{MartensBickPanaggio2016}. Naturally, such networks support three \emph{homogeneous synchronized states}, a fully synchronized state $\SSz = \sset{(R_1,R_2,\psi)=(1,1,0)}$ where both populations are synchronized and in phase, a cluster state $\SSpi = \sset{(R_1,R_2,\psi)=(1,1,\pi)}$ where both populations are synchronized, and in anti-phase and a completely incoherent state $\I = \sset{(R_1,R_2,\psi)=(0,0,*)}$. A bifurcation analysis shows that only one of the three is stable for any given choice of coupling parameters~\cite{MartensBickPanaggio2016}.

In addition to homogeneous synchronized states, networks of two identical populations also support synchronization patterns where synchrony is {localized in one of the two populations}, a chimera, as illustrated in Fig.~\ref{fig:PATTERNSM2}(b). As discussed by Abrams {\it et al.}~\cite{Abrams2008}, for \emph{homogeneous phase-lags} ($\alpha_s=\alpha_n$) stable complete synchrony~$\SSz$ and a stable chimera in $\DS = \sset{R_1<1, R_2=1}$, which is either stationary or oscillatory, coexist\footnote{By symmetry there is a corresponding pattern in~$\SD=\sset{R_1=1, R_2<1}$.}. Note that the Ott--Antonsen reduction simplifies the analysis tremendously: It translates the problem for large oscillator networks into a low-dimensional bifurcation problem. Martens \emph{et al.}~\cite{MartensPanaggioAbrams2016} outlined the basins of attraction of the coexisting stable synchrony patterns and thereby answering the question as to which (macroscopic or microscopic) initial conditions converge to either state.
Through directed perturbations it is possible to switch between different synchrony patterns and thus \emph{functional configurations of the network} that are of relevance in neuroscience~\cite{Kirst2016,Palmigiano2017}, thus embodying memory states or controlling the predominant direction of \emph{information flow} between subpopulations of oscillators~\cite{DeschleMartens2019}. 
Further work addresses the robustness of chimeras against various inhomogeneities, including heterogeneous frequencies~\cite{Laing2009,Laing2012}, network heterogeneity~\cite{Laing2012a}, and additive noise~\cite{Laing2012}.

If one allows for \emph{heterogeneous phase-lag parameters}, $\alpha_s\neq\alpha_n$, a variety of other attractors with localized synchrony emerge~\cite{MartensBickPanaggio2016,Choe2016}. This includes in particular solutions in $\DD=\sset{0<R_1<R_2, R_2<1}$ where neither population is fully phase synchronized; cf.~Fig.~\ref{fig:PATTERNSM2}(b). This includes not only stationary or oscillatory solutions of the state variables, but also attractors where the order parameters $Z_1, Z_2$ fluctuate chaotically both in amplitude and with respect to their phase difference~\cite{Bick2016a}. Finite networks with two populations of identical oscillators may be analyzed using the Watanabe--Strogatz equations~\eqref{eq:WS_KS}. One finds that the bifurcation scenarios for the appearance of chimera states is similar to the dynamics observed for infinite populations~\cite{Panaggio2015b}. Moreover, macroscopic chaos also appears in many finite networks~\cite{Bick2016a} down to just two oscillators per population.

\paragraph{A note on finite networks of identical oscillators and localized frequency synchrony.}
For finite oscillator networks, the widely used intuitive definition of a chimera as a solution for networks of (almost) identical oscillators where ``coherence and incoherence coexist'' is difficult to apply in a mathematically rigorous way. Hence, Ashwin and Burylko~\cite{Ashwin2014a} introduced the concept of a \emph{weak chimera} which provides a mathematically testable definition of a chimera state in finite networks of identical oscillators; here, we only give an intuition and refer to~\cite{Ashwin2014a,Bick2015c} for a precise definition. The main feature of a weak chimera is that identical oscillatory units (with the same intrinsic frequency if uncoupled) generate rhythms with two or more distinct frequencies solely through the network interactions---this is a fairly general form of synchronization. In the context of dynamical systems with symmetry~\cite{Golubitsky2002}, weak chimeras are, as outlined in~\cite{Bick2015d}, an example of dynamical symmetry breaking where identical elements have nonidentical dynamics since their frequencies are distinct.

More specifically, a weak chimera is characterized by localized frequency synchrony in a network of identical oscillators. Similar to the definition of identical populations further above, we say that the oscillators are identical if for a pair of oscillators~$(\sigma,k)$ and~$(\tau, j)$ there exists an invertible transformation of the oscillator indices which keeps the equations of motion invariant. In other words, all oscillators are effectively equivalent. Now $\dot\theta_{\sigma,k}(t)$ is the \emph{instantaneous frequency} of oscillator $(\sigma,k)$---the change of phase at time~$t$---and thus the \emph{asymptotic average frequency} of oscillators~$(\sigma,k)$ is
\begin{equation}\label{eq:AvgFreq}
\Omega_{\sigma,k} = \lim_{T\to\infty}\frac{1}{T}\int_{0}^{T}\dot\theta_{\sigma,k}(t)\udi t.
\end{equation}
Rather than looking at phase synchrony ($\theta_{\sigma,k}=\theta_{\tau, j}$) of oscillators~$(\sigma,k)$ and~$(\tau, j)$, we say that the oscillators are \emph{frequency synchronized} if $\Omega_{\sigma,k}=\Omega_{\tau, j}$. Weak chimeras now show \emph{localized frequency synchrony}, that is, all oscillators within one population have the same frequency $\Omega_{\sigma}=\Omega_{\sigma,k}$ while there are at least two distinct populations $\tau \neq \tau'$ that have different frequencies, $\Omega_{\tau}\neq\Omega_{\tau'}$. Note that weak chimeras are impossible for a globally coupled network of identical phase oscillators (that is, there is only a single population $\maxpop=1$): Such a network structure forces frequency synchrony of all oscillators~\cite{Ashwin2014a}.

Weak chimeras have been shown to exist in a range of networks which consist of $\maxpop=2$ interacting populations of phase oscillators. For weakly interacting populations of phase oscillators with general interaction functions there can be stable weak chimeras with quasiperiodic~\cite{Ashwin2014a,Bick2017} and chaotic dynamics~\cite{Bick2015c}. However, neither weak interaction nor general coupling functions are necessary for dynamics with localized frequency to arise: Even sinusoidally coupled networks~\eqref{eq:PopKuramotoSakaguchi} of just $\maxdim=2$ oscillators per population support stable regular~\cite{Panaggio2015b} and chaotic~\cite{Bick2016a} weak chimeras.

\paragraph{Dynamics of nonidentical populations with distinct frequency distributions.}
As mentioned above, chimera states appear for two identical populations of phase oscillators. Using the Ott--Antonsen equations, Laing showed that these dynamics persist for~\eqref{eq:PopKuramotoSakaguchi} with $\maxpop=2$ if $\Delta_\sigma>0$ and $\omega_\sigma\neq\omega_{\sigma'}$ in the large~$\maxdim$ limit~\cite{Laing2009}; see also~\cite{Skardal2019} for further bifurcation analysis. As heterogeneity is increased, stationary chimera states can become oscillatory through  Hopf bifurcations and may eventually be entirely destroyed.

Montbri\'o {\it et al.}~\cite{Montbrio2004} studied two populations where not only frequencies were nonidentical ($\Delta_\sigma>0$, $\Omega_\sigma\neq \Omega_{\sigma'}$), but also the coupling was asymmetric between the two populations. In another study, Laing \emph{et al.}~considered noncomplete networks to study the sensitivity of chimera states against gradual removal of random links starting from a complete network~\cite{Laing2012a}, and found that oscillations of chimera states can be either created or suppressed depending on the type of link removal.

\paragraph{Dynamics of nonidentical populations with asymmetric input or output.}
Another way to break symmetry in a population of Kuramoto oscillators is inspired by neural networks with excitatory and inhibitory coupling~\cite{borgers2003}: One replaces~$K$ with a random coefficient~$K_j$ \emph{inside} the sum in~\eqref{eq:Kuramoto}. Thus, oscillators with $K_j >0$ mimic the behavior of excitatory neurons while those with $K_j<0$ correspond to inhibitory neurons. The interactions between oscillators~$j$ and~$l$ are not necessarily symmetric, unless $K_j = K_{l}$. The study by Hong and Strogatz~\cite{Hong2012} reveals that---somewhat surprisingly---extending the Kuramoto model in this fashion yields dynamics that resembles that of the original model~\eqref{eq:Kuramoto} when the intrinsic frequencies~$\omega_k$ are nonidentical. 
Similar coupling schemes accommodating for excitatory and inhibitory coupling have been devised for multi-population models~\eqref{eq:KuramotoOPpopulation}, to study how solitary states emerge within a synchronized population, thus leading to the formation of clusters~\cite{Maistrenko2014}.

Another possibility to include coupling heterogeneity considered by Hong and Strogatz is to introduce an oscillator dependent coupling parameter~$K_k$ \emph{outside} of the sum in Eq.~\eqref{eq:Kuramoto}; see~\cite{Hong2011a}. This relates to social behavior: An oscillator~$k$ is conformist if $K_k>0$ (it wants to synchronize) and contrarian if $K_k<0$. This setup may give rise to complex states where oscillators bunch up in groups with a phase difference of~$\pi$ or move like a traveling wave. A later study found that the system with identical oscillators harbors even more complex dynamics, such as incoherent and other states~\cite{Hong2011}.

\subsubsection{Three and more oscillator populations}\label{sec:3PopPhsLg}

\paragraph{Stable synchrony patterns for three identical populations.}

We first consider identical populations with reciprocal coupling in the sense that $c_{\sigma\tau}=c_{\tau\sigma}$; see~\cite{Martens2010bistable,Martens2010var}. Here the coupling is determined by {self}-coupling~$\ks$ and phase-lag~$\as$, as well as coupling strength and phase lag to the {neighboring} populations $k_{n_1},k_{n_2},k_{n_3}$ and $\alpha_{n_1},\alpha_{n_2},\alpha_{n_3}$.
Reducing the phase-shift symmetry, the state of the system is determined by the magnitude of the order parameters, $R_\sigma=\abs{\OP_\sigma}$ and the phase differences between the mean fields $\psi_1=\phi_2-\phi_1$ and $\psi_2=\phi_3-\phi_1$.

Networks of three populations support a variety of localized synchrony patterns. 
For coupling with a {triangular} symmetry, that is, $k_{n_1}=k_{n_2}=k_{n_3}\leq k_s$ and $\alpha_{n_1}=\alpha_{n_2}=\alpha_{n_3}=\alpha_s$,
Martens~\cite{Martens2010bistable} identified coexisitng synchrony patterns: There are three stable solution branches, full phase synchrony $\SSS = \sset{R_1=R_2=R_3=1}$ as well as two chimeras in $\SDS = \sset{R_1=R_3=1>R_2}$ and in $\DSD = \sset{R_1=R_3<R_2=1}$. The Ott--Antonsen reduction allows one to perform an explicit bifurcation analysis of the resulting planar system and shows bifurcations similar to networks with $\maxpop=2$ populations. Remarkably, there are parameter values where~$\SSS$ as well as the chimeras in $\SDS$, $\DSD$ are stable simultaneously; this gives rise to the possibility of switching between these three synchronization patterns through directed perturbations~\cite{MartensPanaggioAbrams2016}.
This triangular symmetry is broken in~\cite{Martens2010var} by allowing $k_{n_2}\neq  k_{n_1}$. Thus, the coupling between populations~2 and~3 can be gradually reduced or increased until the network effectively becomes a chain of three populations or effectively two populations, respectively. A bifurcation analysis shows that the chimeras in~$\SDS$ and~$\DSD$ persist and provides stability boundaries.

\paragraph{Metastability and dynamics of localized synchrony for identical oscillators.}

The synchrony patterns above were primarily considered as attractors: For a range of initial phase configurations, the long term dynamics of the oscillator network will exhibit a particular synchrony pattern. While this may be a good approximation for large scale neural dynamics on a short time-scale, the global dynamics of large-scale brain neural networks are usually much more complicated~\cite{Lehnertz2017}. Neural recordings show that particular dynamical states (of synchrony and activity) persist for some time before a rapid transition to another state~\cite{Abeles95,Britz2010,Tognoli2014}. One approach to model such dynamics is to assume that there are a number of \emph{metastable states} (rather than attractors) in the network phase space which are connected dynamically by \emph{heteroclinic trajectories}\footnote{A heteroclinic trajectory between two distinct saddles is a solution that is attracted to one saddle as time increases and to the other saddle as time evolves backward.}~\cite{AshwinTimme2005}. If heteroclinic trajectories form a \emph{heteroclinic network}\footnote{Unfortunately, the term ``network'' has a double meaning here: on the one hand, we study oscillatory units which form networks through their (physical and functional) interactions, on the other hand, heteroclinic networks are abstract networks of dynamical states linked by heteroclinic trajectories which allow dynamical transitions.}---the nodes of this network are dynamical states, links are connecting heteroclinic trajectories---the system can exhibit sequential switching dynamics: The state will stay close to one metastable state before a rapid transition, or switch, to the next dynamical state. Heteroclinic networks have long been subject to investigations, both theoretically~\cite{Ashwin2017} and with respect to applications in neuroscience~\cite{Ashwin2016}; one possible modeling approach is to write down kinetic (Lotka--Volterra type) equations for interacting macroscopic activity patterns~\cite{Rabinovich2006,Rabinovich2012} which support heteroclinic networks.

Heteroclinic dynamics also arise in phase oscillator networks. For globally coupled oscillator networks, i.e., $\maxpop=1$ population, there are heteroclinic networks between patterns of phase synchrony~\cite{Hansel1993,Ashwin2007}. As mentioned above, all oscillators in these networks are necessarily frequency synchronized, that is, they show the same rate of activity. More recently, it was shown that more general network interactions than those in~\eqref{eq:PopKuramotoSakaguchi} allow for \emph{heteroclinic switching between weak chimeras} as states with localized frequency synchrony~\cite{Bick2017c}: Each population will sequentially switch between states with high activity (frequency) to a state with low activity. One of the simplest phase oscillator networks which exhibits such dynamics consists of $\maxpop=3$ populations of $\maxdim=2$ oscillators where $K>0$ mediates the coupling strength between populations. More precisely, the dynamics of oscillator $(\sigma, k)$ are given by
\begin{equation}
\begin{split}
\dot\theta_{\sigma,k} &= \sin(\theta_{\sigma,3-k}-\theta_{\sigma,k}+\alpha)+r\sin(2(\theta_{\sigma,3-k}-\theta_{\sigma,k}+\alpha))
\\&\qquad
-K\cos(\theta_{\sigma-1,1}-\theta_{\sigma-1,2}+\theta_{\sigma,3-k}-\theta_{\sigma,k}+\alpha)\\&\qquad
-K\cos(\theta_{\sigma-1,2}-\theta_{\sigma-1,1}+\theta_{\sigma,3-k}-\theta_{\sigma,k}+\alpha)\\&\qquad
+K\cos(\theta_{\sigma+1,1}-\theta_{\sigma+1,2}+\theta_{\sigma,3-k}-\theta_{\sigma,k}+\alpha)\\&\qquad
+K\cos(\theta_{\sigma+1,2}-\theta_{\sigma+1,1}+\theta_{\sigma,3-k}-\theta_{\sigma,k}+\alpha).
\end{split}
\end{equation}
Here the interactions within each population is not just given by a first harmonic as in~\eqref{eq:PopKuramotoSakaguchi} but also by a second harmonic (scaled by a parameter~$r$); this is sometimes referred to as Hansel--Mato--Meunier coupling~\cite{Hansel1993}. Moreover, the interactions between populations are not additive but consist of nonlinear functions of four phase variables; this is a concrete example of higher-order interaction terms discussed above. It remains an open question whether such generalized interactions are necessary to generate heteroclinic dynamics between weak chimeras.

Dynamics of metastable states with localized (frequency) synchrony are of interest also in larger networks of $\maxpop>3$ populations. Since explicit analytical results are hard to get for such networks, Shanahan~\cite{Shanahan2010} used numerical measures to analyze how metastable and ``chimera-like'' the network dynamics are. Recall that $R_\sigma(t)$ encodes the level of synchrony of population~$\sigma$ at time~$t$. Let $\avg{\ \cdot\ }_\sigma$, $\Var_\sigma$ denote the mean and variance over all populations $\sigma=1, \dotsc, \maxpop$ and $\avg{\ \cdot\ }_T$, $\Var_T$ mean and variance over the time interval $[0, T]$. Now
\[\lambda = \avg{\Var_T(R_\sigma(t))}_\sigma\]
gives how much synchrony of individual populations vary over time while
\[\chi = \avg{\Var_\sigma(R_\sigma(t))}_T\]
encodes how much synchrony varies across populations. Intuitively, large values of~$\lambda$ correspond to a high level of ``metastability'' while large values of~$\chi$ indicate that the dynamics are ``chimera-like''. On the one hand, these measures have subsequently been applied to more general oscillator networks~\cite{Wildie2012,Deco2017}. On the other hand, they have been applied to study the effect of changes to the network structure (for example through lesions) to the dynamics of Kuramoto--Sakaguchi oscillators~\eqref{eq:PopKuramotoSakaguchi} with delay on human connectome data~\cite{Park2013}.

\paragraph{Populations with distinct intrinsic frequencies.}
Mean-field reductions have also been successful at describing networks of nonidentical populations with distinct mean intrinsic frequencies. Examples of such a setup include interacting neuron populations in the brain with distinct characteristic rhythms.
Resonances between the mean intrinsic frequencies give rise to higher-order interactions. Subject to certain conditions, one can apply the Ott--Antonsen reduction for the mean-field limit~\cite{Komarov2013a} or the Watanabe--Strogatz reduction for finite networks~\cite{Luck2011} to understand the collective dynamics. If resonances between the mean intrinsic frequencies of the populations are absent~\cite{Komarov2011}, then the  mean-field limit equations~\eqref{eq:OttAnt}---a system with $2\maxpop$ real dimensions---simplify even further. More specifically, assume that the intrinsic frequencies are distributed according to a Lorentzian distribution with width~$\Delta_\sigma$ and write $\OP_\sigma = R_\sigma e^{i\phi_\sigma}$ for the Kuramoto order parameter as above. As outlined in~\cite{Komarov2011}, nonresonant interactions imply that---as in~\eqref{eq:KuramotoMFOP}---the equations for~$R_\sigma$ in~\eqref{eq:OttAnt} decouple from the dynamics of the mean phases~$\phi_\sigma$. 
That is, the macroscopic dynamics are described by the $\maxpop$-dimensional system of equations
\begin{equation}\label{eq:NonResOsc}
\dot R_\sigma = \left(-\Delta_\sigma - \sum_{\tau=1}^{\maxpop}b_{\sigma\tau}R_\tau + \left(1-R_\sigma^2\right)\bigg(a_\sigma+\sum_{\tau=1}^{\maxpop}c_{\sigma\tau}R_\tau\bigg)\right)R_\sigma
\end{equation}
where $a_\sigma, b_{\sigma\tau}, c_{\sigma\tau}\in\R$ are parameters which depend on the underlying nonlinear oscillator system. Note that these equations of motion are similar to Lotka--Volterra type dynamical systems which have been used to model sequential dynamics in neuroscience~\cite{Rabinovich2006,Rabinovich2012}. Indeed, \eqref{eq:NonResOsc} give rise to a range of dynamical behavior including sequential localized synchronization and desynchronization through cycles of heteroclinic trajectories and chaotic dynamics~\cite{Komarov2011}.

\subsection{Networks of neuronal oscillators}

Neurons can be modeled at different levels of realism and complexity~\cite{koch2004}. The approach we (and many others) take is to ignore the spatial extent of individual neurons (including dendrites, soma, and axons) and treat each neuron as a single point whose state is described by a small number of variables such as intracellular voltage and the concentrations of certain ions. We also ignore stochastic effects and describe the dynamics of single neurons by a small number of ordinary differential equations. By definition, the state of a Theta neuron or a QIF neuron is described by a phase variable. However, under the assumption of weak coupling, higher-dimensional models with a stable limit cycle (e.g., Hodgkin--Huxley, FitzHugh--Nagumo) can be reduced to a phase description using phase reduction~\cite{Ashwin2016,Pietras2019}.

The two main types of coupling between neurons are through synapses or gap junctions. In synaptic coupling, the firing of a presynaptic neuron causes a change in the membrane conductance
of the postsynaptic neuron, mediated by the release of neurotransmitters. This has the effect
of causing a current to flow into the postsynaptic neuron, the current being of the form
\be
 I(t)=\gt(t)(\Vrev-V), \label{eq:Isyn}
\ee
where~$\Vrev$ is the reversal potential for that synapse, $V$ is the voltage of the postsynaptic neuron, and $\gt(t)$ is the time-dependent conductance. The sign of~$\Vrev$ relative to the resting potential of the postsynaptic neuron governs whether the synapse is excitatory or inhibitory. The function $\gt(t)$ may be stereotypical, i.e.,~it may have
the same functional form for each firing of the presynaptic neuron, where~$t$ is measured from the last firing, or it may have its own dynamics. One approximation in this type of modeling is to ignore the value of~$V$ in~\eqref{eq:Isyn} and just assume that the firing of a presynaptic neuron causes a pulse of current to be injected into the 
postsynaptic neuron(s).

In gap junctional coupling a current flows that is proportional to voltage differences, so if neurons~$k$ and~$j$ have voltages~$V_k$ and~$V_j$, respectively, and~$\gt$ is the (constant) gap junction conductance, the current flowing from neuron~$k$ to neuron~$j$ is $I=\gt(V_k-V_j)$.

\subsubsection{Populations of Theta neurons}
\label{sec:ThetaNeurons}

In this section, we consider a population of Theta neurons~\eqref{eq:dthdtD} where the network interactions are generated by the input from all other neurons in the network. For input through synapses, for example, each neuron receives signals from the rest of the network through the input current~$I$. Here, we will focus on the Ott--Antonsen reduction for Theta neurons~\eqref{eq:dzdtA} in the mean-field limit, assuming that variations in excitability are distributed according to a Lorentzian. The key ingredient here is to write the network input in terms of the mean-field variables to obtain a closed system of mean-field equations; as we will see below, this is possible for a range of couplings that are relevant for neural dynamics. For now, we focus on one population and omit the population index~$\sigma$.

\begin{figure}
\begin{center}
\includegraphics[width=0.6\textwidth]{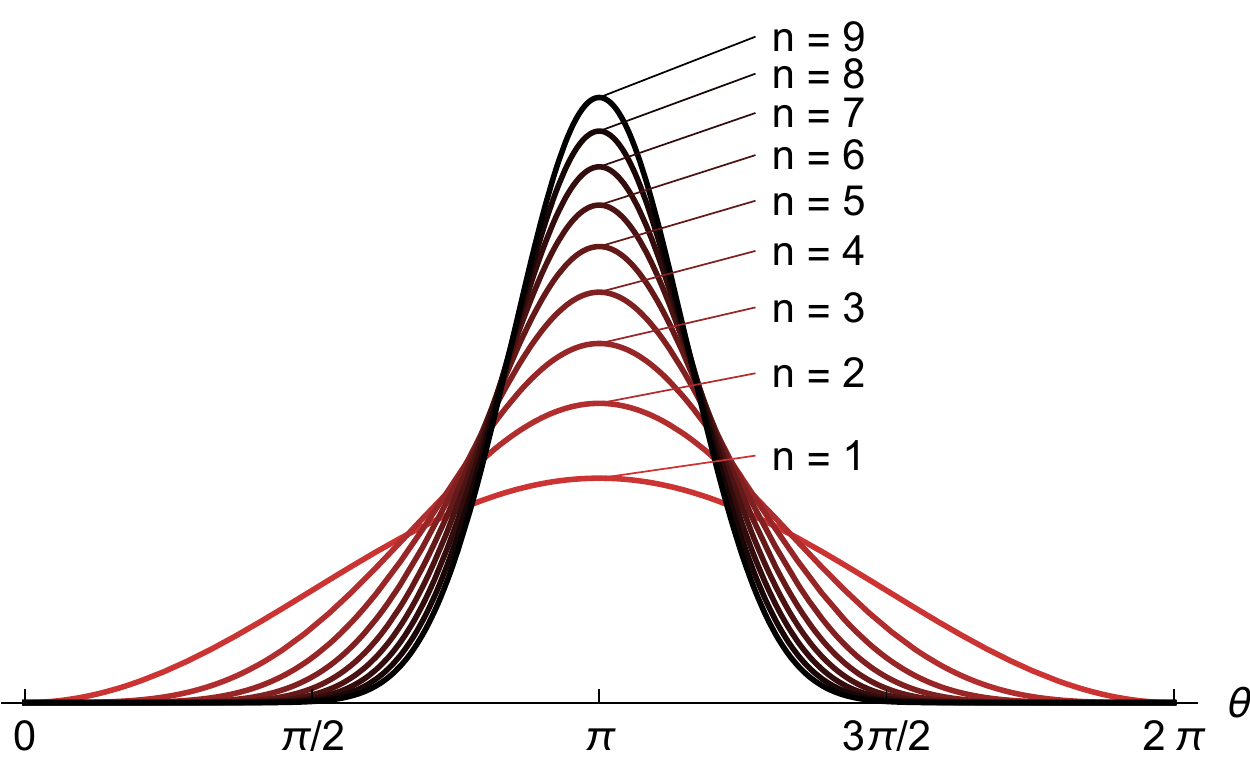}
\smallskip
\end{center}
\caption{\label{fig:Pulse}The function~$P_n(\theta)$ is a pulse centered at $\theta=\pi$, the phase where a neuron spikes; here~$P_n$ is plotted for $n=1, \dotsc, 9$. As~$n$ increases, the pulse becomes narrower.}
\end{figure}

In the following, we consider a network where each neuron emits a pulse-like signal of the form 
\begin{align}\label{eq:Pulse}
P_n(\theta) &= a_n(1-\cos{\theta})^n
\end{align}
as it fires (the phase~$\theta$ increases through~$\pi$, see Figs.~\ref{fig:Pulse} and~\ref{fig:ThetaNeuronSNIC}). The parameter $n\in\N$ determines the sharpness of a pulse and~$a_n = 2^{n} (n!)^2/(2n)!$ is the normalization constant such that $\int_0^{2\pi}P_n(\theta)\udi\theta=2\pi$; cf.~Fig.~\ref{fig:Pulse}.
The average output of all neurons in the network, each one contributing identically, is
\be
   \Pn=\frac{1}{\maxdim}\sum_{j=1}^\maxdim P_n(\theta_j).
\ee
Now~$\Pn$ can be expressed as a function of the order parameter~$\OP$: As shown in~\cite{Luke2013,so2014,laing2016} we have for the mean-field limit of infinitely many neurons, $\maxdim\to\infty$,
\begin{align}\label{eq:PulseZ}
   \Pn&=a_n\left(C_0+\sum_{q=1}^nC_q(\OP^q+\bar{\OP}^q)\right)
   \intertext{with coefficients}
   C_q&=\sum_{k=0}^n\sum_{m=0}^k\frac{n!(-1)^k\delta_{k-2m,q}}{2^k(n-k)!m!(k-m)!}.
\end{align}
Here $\delta_{p,q}=1$ if $p=q$ and $\delta_{p,q}=0$ otherwise. In the limit of infinitely narrow pulses, $n\rightarrow\infty$, we find 
\be\label{eq:PulseZInf}
P^\infty=\frac{1-\abs{\OP}^2}{1+\OP+\bar{\OP}+\abs{\OP}^2}.
\ee

\paragraph{Synaptic coupling.}
If each Theta neuron~\eqref{eq:dthdtD} receives instantaneous synaptic input in the form of current pulses as in~\cite{Luke2013,so2014,Laing2014}, the input current to each neuron is the network output
\be \label{eq:SynapticCplng}
I(t) = \Pn(t).
\ee
A positive coupling strength~$\kappa>0$ for the Theta neuron~\eqref{eq:dthdtD} corresponds to excitatory coupling and $\kappa<0$ to inhibitory coupling. Note that since~$I$ now is a function of the Kuramoto order parameter by~\eqref{eq:PulseZ}, we have closed the Ott--Antonsen equation for the Theta neuron~\eqref{eq:dzdtA} to obtain a system describing the dynamics for infinitely many oscillators.

\medskip
\noindent\textbf{Example~2:\ }\label{example2}%
The challenge in \hyperref[problem2]{\emph{Problem~2}} was to classify what dynamics are possible in a single population of globally coupled Theta neurons with pulsatile coupling and specify the onset where neurons start to fire. The dynamical repertoire of a population of Theta neurons can be understood using the Ott--Antonsen equations for the limit $N\rightarrow\infty$. We follow the work by Luke \emph{et~al.}~\cite{Luke2013} who considered a network with pulsatile coupling~\eqref{eq:Pulse} with nontrivial width $n=2$ and direct synaptic coupling. According to~\eqref{eq:PulseZ} the pulse shape evaluates to 
\be\nonumber
P^{(2)}(Z) = 1 + \frac{1}{6}\OP^2 + \frac{1}{6}\bar{\OP}^2 - \frac{4}{3}\Re{\OP}
\ee
as a function of~$Z$. With direct synaptic coupling~$I=\kappa P^{(n)}$, the Ott--Antonsen equations~\eqref{eq:dzdtA} for an infinitely large population is thus given by
\begin{align}\nonumber
    \dot{\OP} &= -\half\left(\left(\Delta-i\eta - i\kappa\Big(1 + \frac{1}{6}\OP^2 + \frac{1}{6}\bar{\OP}^2 - \frac{4}{3}\Re{\OP}\Big)\right) (1+Z)^2 +i(1-Z)^2\right).
\end{align}
This closed, two-dimensional set of equations can readily be analyzed using dynamical systems methods.

Different dynamic behaviors may be observed for~\eqref{eq:dzdtA} while varying up to three parameters: the coupling strength~$\kappa$, excitability threshold~$\etb$, and the width of their distribution~$\Delta$. Luke \emph{et~al.}~\cite{Luke2013} found three distinct stable dynamical regimes: (i)~\emph{partially synchronous rest}, (ii)~\emph{partially synchronous spiking}, and (iii)~\emph{collective periodic wave} dynamics. 
In partially synchronous rest, most neurons remain quiescent (a stable node in the two-dimensional Ott--Antonsen equations~\eqref{eq:dzdtA} for~$\OP$); in the partially synchronous spiking regime most neurons spike continuously (a stable focus for~$\OP$); and in the collective periodic wave neurons fire periodically (a stable periodic orbit of the order parameter~$\OP$). 
Varying~$\kappa$ from small to large values, we typically observe a transition from partially synchronous spiking (quiescence) to partially synchronous spiking, with growing synchrony as~$\kappa$ increases. This transition is characterized by hysteresis arising around two fold bifurcations, originating in a cusp bifurcation. For certain parameter values, the order parameter may undergo a Hopf bifurcation from partially synchronous spiking to collective periodic wave dynamics so that~$\OP(t)$ becomes oscillatory.

In a neuroscientific context, knowledge of macroscopic quantities other than the order parameter~$\OP$ is sometimes preferable, such as the firing rate given via~\eqref{eq:firingrate1} as $r=\frac{1}{\pi}\Re{(1-\bar{\OP})/(1+\bar{\OP})}$. Alternatively, as outlined in Section~\ref{sec:FRE}, the macroscopic equation~\eqref{eq:dwdt} for~$W(t)$ is equivalent to \eqref{eq:dzdtA} via  a conformal transformation and describes the evolution of the population's firing rate $r=\frac{1}{\pi}\Re{W}$ and average voltage $V=\Im{W}$. The algebraic solution for stationary states is particularly simple if one chooses infinitely narrow pulse shape ($n\rightarrow \infty$); however, note that this choice may be biophysically less realistic~\cite{Luke2013} and renders more degenerate dynamic behavior, e.g., bifurcations giving rise to oscillations in the order parameter (firing rate) disappear in this particular network with $M=1$ population. 

Finally, we note that if in addition the excitability of neurons varies periodically, more complicated dynamics and macroscopic chaos can be observed~\cite{so2014}. While this example covers networks of Theta neurons, the same approach applies to networks with QIF neurons with direct synaptic coupling as given by~\eqref{eq:SynapticCplng}; see, for example, the analyses in~\cite{Devalle2017,Ratas2017,Ceni2019}.

\medskip

A simple modification of~\eqref{eq:dthdtD} is to add synaptic dynamics by letting the input current~$I$ satisfy the equation
\be\label{eq:SynapticDyn}
   \ts\dot I=\Pn-I,
\ee
where $\ts$ is the time-constant governing the synaptic dynamics. In the limit $\ts\to 0$ the synaptic dynamics are instantaneous and we recover the previous model. Again, with~\eqref{eq:PulseZ} the Ott--Antonsen equations~\eqref{eq:dzdtA} and~\eqref{eq:SynapticDyn} form a closed system of equations that describe the dynamics in the mean-field limit.

\paragraph{Gap junctions.} Along with synaptic coupling, the other major form of coupling between neurons is via gap junctions~\cite{coombes2008}, in which a current flows between connected neurons proportional to the difference in their voltages. Using the equivalence of the Theta and QIF neuron, it was shown in~\cite{laing2015} that
adding all-to-all gap junction coupling to~\eqref{eq:dthdtD} results in the equations
\be
   \dot\theta_k=1-\cos{\theta_k}-\kgj\sin{\theta_k}+(1+\cos{\theta_k})\left(\eta_k+\kappa I+\frac{\kgj}{N}\sum_{j=1}^N\tn(\theta_j)\right) \label{eq:dthdtB},
\ee
where~$\kgj$ is the strength of gap junction coupling and the function $\tn(\theta):=\sin{\theta}/(1+\cos{\theta}+\epsilon)$ with $0<\epsilon\ll 1$ stems from the coordinate transformation between Theta and QIF neurons. Note that~\eqref{eq:dthdtB} is still a sinusoidally coupled system. Assuming a Lorentzian distribution of excitability~$\eta_k$ centered at~$\etb$ with width~$\Delta$, the dynamics in the limit of infinitely many oscillators are given by the Ott--Antonsen equation
\be
   \dot \OP=\frac{1}{2}\left((i\etb-\Delta)(1+\OP)^2-i(1-\OP)^2\right)+\frac{1}{2}\left(i(1+\OP)^2(\kappa I+\kgj Q)+\kgj(1-\OP^2)\right) \label{eq:dzdtB},
\ee
where
\begin{align}
   Q&=\sum_{m=1}^\infty \left(b_m \OP^m+\bar b_m \bar \OP^m\right), & b_m&=\frac{i(\rho^{m+1}-\rho^{m-1})}{2\sqrt{2\epsilon+\epsilon^2}},
   \end{align}
and $\rho=\sqrt{2\epsilon+\epsilon^2}-1-\epsilon$. Note that the input current is still to be defined: There could be gap junction only coupling, $I=0$, instantaneous synaptic input~\eqref{eq:SynapticCplng} or synaptic dynamics~\eqref{eq:SynapticDyn} as defined above.

The reduced equations allow one, for example, to study what effect the strength of the gap junction coupling has on the dynamics. Laing~\cite{laing2015} found that for excitatory synaptic coupling (i.e.,~$\kappa>0$) increasing the strength of gap junction coupling could induce oscillations in the mean field via a Hopf bifurcation, and destroy previously existing bistability between steady states with high and low mean firing rates. For inhibitory synaptic coupling (i.e.,~$\kappa<0$) increasing the strength of gap junction coupling stabilized a steady state with high mean firing rate, inducing bistability in the network. In spatially extended systems, it was found that gap junction coupling
could destabilize ``bump'' states via a Hopf bifurcation, and create traveling waves of activity.

Note that in recent work~\cite{PietrasDevalle2019} the authors showed that one can take the limit $\epsilon\rightarrow 0$ in the above derivation, thus simplifying the analysis and allowing one to treat synaptic and gap junctional coupling (in an infinite network of QIF neurons) on equal footing.

\paragraph{Conductance dynamics.} The above models for Theta neurons have all assumed that
synaptic coupling is via the injection of current pulses. However, Ref.~\cite{Coombes2016} considers a model in which synaptic input was in the form of a current, equal to the product of a conductance and the difference between the voltage of a QIF neuron and a reversal potential~$\Vrev$. Converting to Theta neuron variables, a particular case of their model can be written as
\be
   \dot\theta_k = 1-\cos{\theta_k}+(1+\cos{\theta_k})\left(\eta_k+\gt(t)\Vrev\right)-\gt(t)\sin{\theta_k} \label{eq:dthdtC}
\ee
with a time-dependent gating function
\be\label{eq:Gating}
   \gt(t)=\kg P^{\infty}(t)
\ee
that depends on the network output modulated by the coupling strength~$\kg>0$.
(Note that quantities like~$\gt$ and~$\Vrev$ have been non-dimensionalized by scaling relative
to  dimensional quantities.) 
The corresponding Ott--Antonsen equations read
\be
   \dot \OP=\frac{1}{2}\left((i\etb-\Delta)(1+\OP)^2-i(1-\OP)^2\right)+\frac{1}{2}\left(i(1+\OP)^2\gt\Vrev+(1-\OP^2)\gt\right) \label{eq:dzdtC}
\ee
which is closed since $\gt(t)$ is a function of~$\OP$ by~\eqref{eq:PulseZInf}.

The dynamics of this network are straightforward and as expected: For inhibitory coupling
($\Vrev<0$) there is one stable fixed point for all~$\etb$ while for excitatory coupling
($\Vrev>0$) there can be a range of negative~$\etb$ values for which the network
is bistable between steady states with high and low average firing rates. This bistability in an excitatorially self-coupled network is of interest as such a network can be thought of as 
a one-bit ``memory'', stably storing one of two states.

\subsubsection{Populations of Winfree oscillators}
The state of a Winfree oscillator~\cite{winfree1967} is also described by a single angular variable. The Winfree model predates the Kuramoto model and mimics the behavior of biological systems such as flashing fireflies or circadian rhythms in Drosophila~\cite{ariaratnam2001}. In general, the Winfree model does not exhibit sinusoidal coupling. But under suitable assumptions, a network of Winfree oscillators is amenable to simplification through the Ott--Antonsen reduction~\cite{Pazo2014}. Consider a network of~$\maxdim$ Winfree phase oscillators which evolve according to
\be
   \dot\theta_k=\omega_k+\frac{\epsilon}{N}\sum_{j=1}^\maxdim \hat{P}(\theta_j)Q(\theta_k)
\ee
for $k=1, \dotsc, \maxdim$ and $2\pi$-periodic functions~$Q$ and $\hat{P}$. The function~$Q$ is the phase response curve of an oscillator, which can be measured experimentally or determined from a model neuron~\cite{schultheiss2011}. If we set $Q(\theta)=\sin{\beta}-\sin{(\theta+\beta)}$ with parameter~$\beta$ then we have a sinusoidally coupled phase oscillator network. Moreover, suppose that network interaction is given by a pulsatile function~$\hat{P}(\theta)=P_n(\theta-\pi)$. While~$\hat{P}$ has its maximum at $\theta=0$ (unlike the interactions for the Theta neuron), it can be expanded in a similar way as~\eqref{eq:PulseZ} into powers of the Kuramoto order parameter. Assuming that the intrinsic frequencies are distributed as a Lorentzian, we obtain an Ott--Antonsen equation that describes the dynamics in the limit of infinitely large networks; see~\cite{Pazo2014} for details.

Several groups have used this description to study the dynamics of infinite networks
of Winfree oscillators. Paz\'o and Montbri\'o~\cite{Pazo2014} found that such a network typically has either an
asynchronous state (constant mean field) or a synchronous state (periodic oscillations
in the mean field, indicating partial synchrony within the network) as attractors.
They also found that varying~$n$ (the sharpness of~$P_n$) had a significant effect
on the synchronizability of the network. Laing~\cite{laing2016} studied a spatially-extended
network of Winfree oscillators and found a variety of stationary, traveling, and chaotic
spatiotemporal patterns. Finally, Gallego~\emph{et al.}~\cite{gallego2017} extended the work in~\cite{Pazo2014}, considering
a variety of types of pulsatile functions and phase response curves.

\subsubsection{Coupled populations of neurons}
While the previous sections discussed a network consisting of a single population of all-to-all coupled model neurons, an obvious generalization is to consider networks of two or more populations. Consider~$\maxpop$ populations of Theta neurons and let $\Pn_\tau$ denote the output of population~$\tau$. For example for synaptic interaction amongst populations, \eqref{eq:SynapticCplng} generalizes to
\be
I_\sigma(t) = \sum_{\tau=1}^\maxpop \kappa_{\sigma\tau}\Pn_\tau(t),
\ee
where $\kappa_{\sigma\tau}$ is the input strength from population~$\tau$ to population~$\sigma$. Writing each~$P_\tau$ in terms of the order parameter~$\OP_\tau$ of population~$\tau$, we obtain a closed set of~$\maxpop$ Ott--Antonsen equations~\eqref{eq:dzdtA} that describe the dynamics for infinitely large populations.

Interacting populations of neural oscillators give rise to neural rhythms. Laing~\cite{laing2016} considered a network of two coupled populations of Theta neurons, one inhibitory and one excitatory. Such networks support a periodic PING rhythm~\cite{borgers2003} in which the activity of both populations is periodic, with the peak activity of the excitatory population activity preceding that of the inhibitory one. Analyses of similar types of networks were performed in~\cite{luke2014,Montbrio2015,Coombes2016}. Periodic behavior of the mean-field equations of coupled populations of Theta neurons (or equivalently QIF neurons) allows one to extract macroscopic phase response curves~\cite{Dumont2017} which allows one to treat such ensembles as single oscillatory units in weakly coupled networks.

Coupled populations of Winfree oscillators support a range of dynamics. In Ref.~\cite{Pazo2014} the authors considered a symmetric pair of networks of Winfree oscillators. They observed a variety of dynamics such a quasiperiodic chimera state in which one population is perfectly synchronous while the order parameter of the other undergoes quasiperiodic oscillations. They also found a chaotic chimera state where one population is phase synchronized while the order parameter of the other one fluctuates chaotically.

\subsubsection{Further generalizations}
The oscillator populations considered above do not have any sense of space themselves, apart from possibly two networks being at different points in space. The brain is three-dimensional, although the presence of layered structures could lend itself to a description in terms of a series of coupled two-dimensional domains. Regardless, the spatial aspects of neural dynamics should not be ignored. Several authors have generalized the techniques discussed above to spatial domains, deriving \emph{neural field} models: spatiotemporal evolution equations for macroscopic quantities~\cite{laing2016a,laing2015,laing2016,Laing2014,esnaola2017,Byrne2019a,laiome20}. The main advantage of using this new generation of neural field models is that unlike classical models~\cite{wilson1973,amari1977}, the derivations from networks of Theta neurons are exact rather than heuristic. Rather than considering neural field models on continuous spatial domains, one could consider them on a discretized network where each node is a brain region and coupling strength are given, for example, by connectome data. We will briefly touch upon these approaches in Section~\ref{sec:NeuroModel} below.

All of the networks above have been all-to-all coupled which is rarely the case in real-world systems. The in-degree of a neuron is the number of neurons connecting to it, whereas the out-degree is the number of neurons to which it connects. For all-to-all coupled networks all neurons have the same in- and out-degree ($\maxdim-1$ for a network of~$\maxdim$ neurons with no self-coupling). Several groups have considered networks in which the degrees are distributed, having a power law distribution, for example~\cite{chandra2017,Laing2012a,blasche2020,laing2020}. The mean-field reduction techniques discussed above can be used to accurately and efficiently investigate the influence of this aspect of network structure on dynamics, and this is of great interest.

Networks of identical oscillators (whether finite or infinite) are described by the Watanabe--Strogatz equations. While the application to Kuramoto-type oscillator networks is fairly standard, the corresponding mean-field equations for Theta neurons~\eqref{eq:WSth} have only recently been analyzed.

\section{Applications to neural modeling}
\label{sec:NeuroModel}

The mean-field reductions and their applications to populations of neural units---as next-generation neural mass models---can give new modeling approaches to understand the dynamics of large-scale neural networks. In the previous section, we took a descriptive dynamical systems perspective to understand the asymptotic dynamics and their bifurcations. We now change the perspective to elucidate how the mean-field reductions can give new insights into neural network dynamics.

\subsection{Dynamics of neural circuits and populations}

\noindent\textbf{Example~3:\ }\label{example3}
How does a heterogeneous network of all-to-all coupled QIF neurons react to a transient stimulus (see \hyperref[problem3]{\emph{Problem~3}} above)? To answer this question using exact mean-field reductions, we analyze a situation similar to that studied by Montbri\'o \emph{et al.}~\cite{Montbrio2015}: Consider a network of QIF neurons~\eqref{eq:QIF} with dynamics governed by
\be
  \dot{V}_k=V_k^2+I_k+\kappa r(t)+s(t) \label{eq:QIFnet}
\ee
for $k=1,\dotsc, N$ with the rule that when the voltage $V_k=\infty$, it is reset to $V_k=-\infty$. The~$I_k$ are chosen from a Lorentzian distribution with mean~$\widehat{\eta}$ and width parameter~$\Delta$, and neurons are coupled all-to-all with coupling strength~$\kappa$. The mean firing rate is given by
\[
  r(t)=\frac{1}{N}\sum_{j=1}^N\sum_\ell\delta(t-t_j^\ell),
\]
representing the average neural activity in the past, i.e., ~$t_j^\ell$ is the firing time of the~$j$th neuron and~$\ell$ is summed only over past firing times, $t_j^\ell<t$. The input current~$s(t)$ will be specified below. Letting $N\rightarrow\infty$ the network is described by~\eqref{eq:dwdt}, which in the present case becomes 
\be
   \dot{W}=i\widehat{\eta}+\Delta-iW^2+i\left(\kappa r(t)+s(t)\right) ,\label{eq:QIFm}
\ee
where $r=\Re{W}/\pi$. Suppose we set $\Delta=0.1,\etb=-0.5$ and $\kappa=5$. Having $\etb<0$ means that in the absence of coupling most neurons will be excitable rather than firing, and $\kappa>0$ models excitatory coupling.

\begin{figure}
\begin{center}
\includegraphics[width=0.9\textwidth]{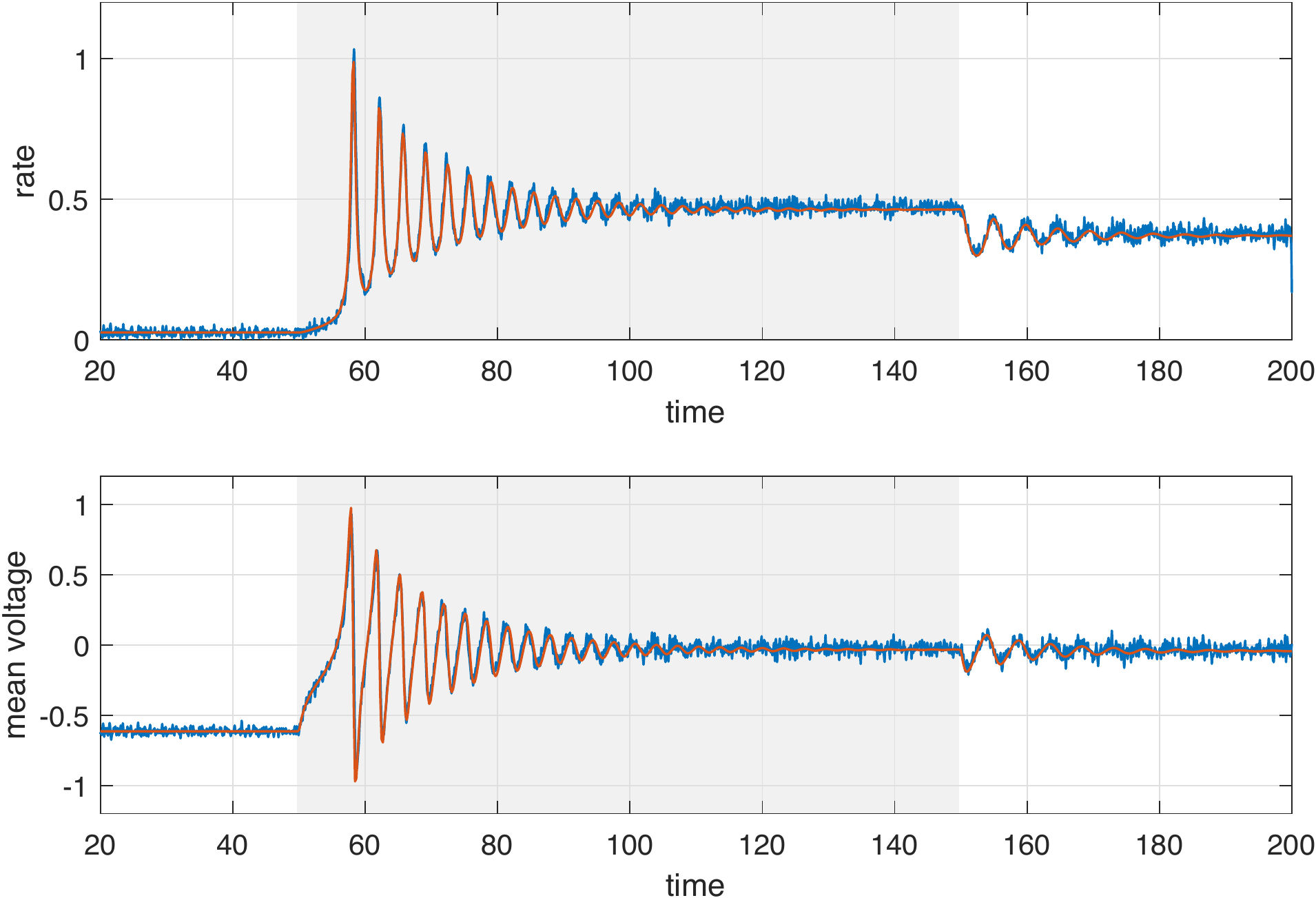}
\smallskip
\end{center}
\caption{\label{fig:QIF}Response of the network~\eqref{eq:QIFnet} (blue lines) and the mean field description~\eqref{eq:QIFm} (red lines) to a transient input turned on at $t=50$ and off at $t=150$ (shaded background).
A network of 1000 neurons was used and the data from the network was smoothed by convolution with a Gaussian of standard deviation 0.05 time units before plotting.
}
\end{figure}

We set the transient input to be $s(t)=0.3$ for $50<t<150$ and $s(t)=0$ otherwise.
The mean firing rate and voltage (i.e., averages over the ensemble of all neurons) of the network are shown in Fig.~\ref{fig:QIF} (top and bottom, respectively) for both the network~\eqref{eq:QIFnet} and the mean field description~\eqref{eq:QIFm}. For these parameters the network is bistable: After the input current is removed, the network settles into an active state rather than returning to the quiescent state that it was in before stimulation. The agreement between the two descriptions is excellent, but the mean-field description is obviously much easier to numerically integrate and is also amenable to bifurcation analysis, as for example shown in~\cite{Montbrio2015}.
\medskip

The influence of oscillatory drive on network dynamics related to cognitive processing in simple working memory and memory recall tasks was studied by Schmidt~{\it et al.}~\cite{Schmidt2018} in coupled populations of inhibitory and excitatory QIF neurons. The authors use the exact mean-field reductions reviewed here to elucidate how oscillatory input frequency stimulates the intrinsic dynamics in networks of recurrently coupled spiking neurons to change memory states. They find that slow delta and theta band oscillations are effective in activating network states associated with memory recall, while faster beta oscillations can serve to clear memory states via resonance mechanisms. 

Balanced sparse networks of inhibitory QIF neurons were studied by di~Volo and Torcini~\cite{diVolo2018} to explain the onset of self-sustained collective oscillations via reduction to mean-field dynamics. This is achieved by applying the mean field reductions to sparse networks with diverging coupling strength, an approximation which works surprisingly well as their bifurcation diagrams show the onset of collective oscillations. The application of the mean-field reductions to sparse networks is ad hoc and further mathematical insights to define a well-defined limit would be desirable.

The work by Dumont and Gutkin~\cite{Dumont2019} used exact phase reductions to identify the biophysical neuronal and synaptic properties that are responsible for macroscopic dynamics, such as the interneuronal gamma (ING) and the pyramidal-interneuronal gamma (PING) population rhythms. The key ingredient is the phase response curve of oscillatory macroscopic behavior of two coupled populations of QIF neurons~\cite{Dumont2017}, one excitatory and one inhibitory, as mentioned above. Assuming weak coupling between two sets of two populations (i.e., four populations total) the authors extracted phase locking patterns of the coupled multipopulation model.

A number of other studies have employed mean-field reductions for populations of QIF neurons to elucidate how microscopic neural properties affect the macroscopic 
dynamics~\cite{bi2020,keeley2019}.
This includes insights into networks of heterogeneous QIF neurons with time delayed, all-to-all synaptic coupling~\cite{Devalle2018,Ratas2018}, or two such networks~\cite{akao2019}.
Moreover, the mean-field reductions are also useful to analyze spatially extended networks of both Theta and QIF neurons, where localized patterns---such as bump states---can occur; cf.~\cite{Laing2014,Byrne2019a}.

\subsection{Large-scale neural dynamics}

The theory above is particularly pertinent for the study of mesoscopic or macroscopic brain dynamics, i.e., dynamics arising from tissue that contains large populations of neurons. Such dynamics are recorded using a variety of different modalities in animal or human studies, including local field potentials (LFP) and magneto- or electroencephapholographic (MEG/EEG) recordings~\cite{Bassett2017}. These recording modalities pick up changes in dynamics that arise in conjunction with fluctuations in populations of neurons. Thus, when recordings are taken from multiple sensors in different positions simultaneously, one can map the spatiotemporal dynamics of large regions of the brain. The inclusion of multiple sensors yields a natural way to construct a large-scale network representation of the dynamics of the brain, in which sensors are nodes of the network. Alternatively, dynamics can be attributed to distributed regions of interest within the brain, for example using approaches to solve the inverse problem and thereby reconstruct a network in source space~\cite{Jonmohamadi2014,Hassan2014}. 

Having defined nodes, to determine interactions~\cite{Stankovski2017} there are several ways to define the edges of large-scale brain networks; in a general context this inverse problem is known as network reconstruction~\cite{Timme2014}. Broadly speaking, edges of brain networks can be characterized as either functional, structural, or effective connections~\cite{Friston2011,Bassett2017}. In the former, a measure of statistical interrelation is used to quantify the extent that the dynamics of nodes co-evolve (see, for example,~\cite{Wang2018}), with edges linking pairs of nodes that are highly correlated being assigned large weights. Structural connectivity, on the other hand, describes a means to define edges on anatomical grounds, for example via tracing of axonal tracts~\cite{Garces2016}. Finally, edges in effective connectivity networks are defined as connection strengths in explicit dynamic models that are tuned such that dynamic recordings are well explained by the model~\cite{Valdes-Sosa2011}.

These different ways of representing the brain in terms of networks yield several avenues for investigation that are relevant to the discussion above. Specifically, network analyses have provided insight into the mechanisms of both function and dysfunction~\cite{Fries2005,Stam2014,Bastos2015,Fornito2015}, and modeling frameworks such as those described above are required in order to explain findings and develop testable predictions~\cite{Bassett2018}. A particularly pertinent question is to understand to what extent structural connectivity---the structural property of the network---shapes emergent functional connectivity---properties of the dynamics---in both healthy and disease conditions~\cite{Honey2007,Honey2010,Senden2014,Bassett2017,Demirtas2017,Misic2016,Shen2015}

Functional connectivity has been shown to be altered in myriad disorders of the brain, including epilepsy and Alzheimer's disease~\cite{Dauwels2010,Lehnertz2014,Stam2014,Schmidt2016,Tait2018}. It is therefore becoming an important marker for brain disorders, as well as a potentially important means of understanding disease and designing therapy~\cite{Goodfellow2016,Fornito2015}. However, in order to link different data modalities and to develop effective and efficient treatment, it is crucial to understand why specific changes in dynamics occur. The reduction methods described herein could help in this direction by bridging fundamental properties of neurons into emergent properties of neuronal networks, which can then be coupled to build an understanding of mesoscopic or whole-brain dynamics~\cite{Breakspear2017}. 

We conclude this section with two very recent examples how the mean-field reductions used here have been used to understand the dynamics of macroscopic brain activities from experimental data. First, Weerasinghe \emph{et al.}~\cite{Weerasinghe2018} employed the Kuramoto model and its mean-field reduction to develop new closed-loop approaches for deep brain stimulation to improve treat patients with essential tremor and Parkinson's disease. Specifically, the Ott--Antonsen equations yield expressions for the mean-field response of an oscillator population, which can be compared with experimentally measured response curves obtained from patients~\cite{Cagnan2017}. The idea is that such a model-supported approach eventually yields efficient treatment strategies, for example, by stimulating at the optimal phase and amplitude to maximize efficacy and minimize side effects.
Second, Byrne \emph{et al.}~\cite{Byrne2019} recently developed a novel brain model based on coupled populations of QIF neurons and use it in a number of neurobiological contexts, such as providing an understanding of the changes in power-spectra observed in EEG/MEG neuroimaging studies of motor-cortex during movement. Such a model is the first step to bridge the microscopic properties of individual neurons to macroscopic brain dynamics.

\section{Conclusions and open problems}

The mean-field descriptions presented in this review are able to bridge spatial scales in coupled oscillator networks since they provide explicit descriptions of the macroscopic dynamics in terms of microscopic quantities. This provides insights into how network coupling properties (for example, a neural connectome) relate to dynamical (and thus functional properties) of an oscillator network. Importantly, the equations are not just a black box, but tools from dynamical systems theory that allow us to study explicitly how the dynamics change as network parameters are varied. We conclude by highlighting three sets of challenges for future research.

The first set of challenges relates to the reductions themselves and the mathematics behind them; some of them were already discussed in Section~\ref{sec:Limitations}, and further along the way. Typically, phase oscillators in the weak coupling limit obey nonsinusoidal coupling, because their interaction contains higher harmonics and nonadditive terms that can arise through strongly nonlinear oscillations or nonlinear interactions between oscillators; see~\cite{Rosenblum2007,Ashwin2015a,Leon2019a} and other references above. Hence, the influence of such interactions on the mean-field reductions still needs to be clarified: While they could fail in certain instances~\cite{Lai2013a}, first results indicate that they may still provide useful information over some timescales~\cite{Vlasov2016}---further work in this direction is desirable.
As an example, Thiem {\it et al.}~\cite{thiem2020} recently used manifold learning and a neural network
to learn the Ott--Antonsen equations governing the Kuramoto model; these techniques are
quite generally applicable.
 Real-world networks are often modeled as systems subject to noise. Here, we point to very recent results that extend the mean-field reductions presented here in these directions by using a ``circular cumulants'' approach~\cite{Tyulkina18,Goldobin2018,Goldobin2018b}.

The second set of challenges concerns the relationship between the mean-field reductions, the underlying microscopic models, and real-world data in the context of neuroscience. How do LFP or EEG measurements relate to the mean-field variables that constitute the reduced system equations? Connectivity can be estimated via neural imaging techniques, but how does this data relate to the coupling strength and phase-lag parameters that appear in the Ott--Antonsen equations of coupled Kuramoto--Sakaguchi populations? Or how does data relate to the coupling parameters of the microscopic models that are compatible with the reduction? These questions become even more intricate for coupled populations of Theta neurons; cf.~\cite{Byrne2019}.

The last set of challenges goes well beyond the mean-field reductions presented here. Mathematical tools are helpful to describe the dynamics, but how do the dynamics relate to functional aspects of the (neural) oscillator network? How do we identify dynamics that are pathological, and validate and use models of these dynamics to predict treatment responses? On the large scale, some pathologies such as epilepsy reveal salient abnormal dynamics~\cite{Kahana2006}, but alterations in other conditions are more subtle, and therefore model-driven analyses could prove itself to be very useful in the clinical context~\cite{Schmidt2016,Kuhlmann2018,Weerasinghe2018,Tait2018}. Insights into these fundamental questions will allow one to make the mean-field reductions presented in this review even more useful to design targeted therapies for neural diseases.

\section*{Acknowledgments}
We are grateful to R.~Bogacz, B.~Duchet, B.~J\"uttner, K.~Lehnertz, and W.~Woldman for helpful feedback on the manuscript, and to R.~Mirollo and J.~Engelbrecht for helpful discussions. We thank the anonymous referees for careful reading and suggestions that helped to improve this review.
This article is part of the research activity (EAM, CL) of the Advanced Study Group 2017 ``From Microscopic to Collective Dynamics in Neural Circuits'' held at Max Planck Institute for the Physics of Complex Systems in Dresden (Germany).

\section*{Funding}
MG gratefully acknowledges the financial support of the EPSRC via grants EP/P021417/1 and EP/N014391/1. MG also acknowledges the generous support of a Wellcome Trust Institutional Strategic Support Award (https://wellcome.ac.uk/) via grant WT105618MA.

\nolinenumbers

\bibliography{refs}

\end{document}